\documentclass[10pt,journal]{IEEEtran} 		
\usepackage[T1]{fontenc} 
\usepackage{setspace} 
\usepackage[
bookmarks=true,
colorlinks=true,
linkcolor=black,
anchorcolor=black,
urlcolor=black,
citecolor=black]{hyperref}
\usepackage{cite}
\usepackage{float}
\usepackage{caption}
\usepackage{subcaption}
\usepackage{graphicx}
\usepackage{tabularx}
\usepackage{booktabs}
\usepackage[table]{xcolor}
\newcolumntype{L}{>{\hspace*{-\tabcolsep}}l}
\newcolumntype{R}{c<{\hspace*{-\tabcolsep}}}
\usepackage{ifthen}
\usepackage{amsmath}
\usepackage{bm} 
\usepackage{amssymb}
\usepackage{mathrsfs}
\usepackage{amsfonts}
\usepackage{commath}
\allowdisplaybreaks[4]
\usepackage{indentfirst}
\usepackage{algorithm,algorithmic}
\usepackage{cleveref}
\newtheorem{myprop}{Proposition}

\newcommand{\modd}{\mathrm{mod}}
\newcommand{\asy}{\mathrm{asy}}
\newcommand{\diag}{\mathrm{diag}}

\newcommand{\trace}{\mathrm{tr}}

\newcommand{\tsce}{\mathrm{tsce}}

\newcommand{\IFFT}{\mathrm{IFFT}}

\newcommand{\cps}{\mathrm{cps}}
\newcommand{\sabs}[1]{\lvert#1\rvert}
\newcommand{\snorm}[1]{\lVert#1\rVert}

\newcommand{\dint}{\,\mathrm{d}}

\newcommand{\ut}{\mathrm{ut}}
\newcommand{\sat}{\mathrm{sat}}

\newcommand{\rc}{\mathrm{c}}

\newcommand{\rd}{\mathrm{d}}
\newcommand{\re}{\mathrm{e}}

\newcommand{\rn}{\mathrm{n}}
\newcommand{\rp}{\mathrm{p}}

\newcommand{\rt}{\mathrm{t}}
\newcommand{\rw}{\mathrm{w}}
\newcommand{\rx}{\mathrm{x}}
\newcommand{\ry}{\mathrm{y}}
\newcommand{\rv}{\mathrm{v}}

\newcommand{\rB}{\mathrm{B}}
\newcommand{\rC}{\mathrm{C}}

\newcommand{\rH}{\mathrm{H}}
\newcommand{\rL}{\mathrm{L}}
\newcommand{\rR}{\mathrm{R}}
\newcommand{\Mx}{M_{\rx}}
\newcommand{\My}{M_{\ry}}

\newcommand{\Npe}{N_{\mathrm{p}}^{\mathrm{e}}}

\newcommand{\Nc}{N_{\mathrm{c}}}
\newcommand{\Nd}{N_{\mathrm{d}}}
\newcommand{\Np}{N_{\mathrm{p}}}
\newcommand{\Ng}{N_{\mathrm{g}}}
\newcommand{\Tc}{T_{\mathrm{c}}}
\newcommand{\Tg}{T_{\mathrm{g}}}

\newcommand{\Ts}{T_{\mathrm{s}}}

\newcommand{\Ld}{L_{\mathrm{d}}}
\newcommand{\mud}{\mu_{\mathrm{d}}}

\newcommand{\FNpNd}{\bdF_{\Np,\Nd}}

\newcommand{\FNpNpe}{\bdF_{\rp,\re}}
\newcommand{\comma}{\text{,}}

\newcommand{\Complex}[2]{\bbC^{#1 \times #2}}
\newcommand{\Real}[2]{\bbR^{#1 \times #2}}

\newcommand{\Xvec}[1]{\mathrm{vec}(#1)}
\newcommand{\Xinv}[1]{\left(#1\right)^{-1}}
\newcommand{\xinv}[1]{\frac{1}{#1}}

\newcommand{\xdeg}[1]{#1^{\circ}}
\newcommand{\xceil}[1]{\lceil#1\rceil}

\newcommand{\bbC}{\mathbb{C}}
\newcommand{\bbE}{\mathbb{E}}
\newcommand{\bbR}{\mathbb{R}}

\newcommand{\clA}{\mathcal{A}}

\newcommand{\clD}{\mathcal{D}}

\newcommand{\clI}{\mathcal{I}}
\newcommand{\clK}{\mathcal{K}}

\newcommand{\clN}{\mathcal{N}}

\newcommand{\clO}{\mathcal{O}}

\newcommand{\clQ}{\mathcal{Q}}

\newcommand{\clW}{\mathcal{W}}
\newcommand{\clX}{\mathcal{X}}

\newcommand{\clCN}{\mathcal{CN}}

\newcommand{\vtheta}{\vartheta}

\newcommand{\bva}{\breve{a}}
\newcommand{\bvbdh}{\breve{\bdh}}
\newcommand{\ckbdh}{\check{\bdh}}

\newcommand{\cka}{\check{a}}

\newcommand{\htbdd}{\hat{\bdd}}

\newcommand{\htbdH}{\hat{\bdH}}

\newcommand{\tdbdx}{\tilde{\bdx}}
\newcommand{\tdbdy}{\tilde{\bdy}}

\newcommand{\udbdU}{\underline{\bdU}}
\newcommand{\udbdW}{\underline{\bdW}}

\newcommand{\bdzro}{\mathbf{0}}

\newcommand{\bda}{\mathbf{a}}
\newcommand{\bdb}{\mathbf{b}}

\newcommand{\bdd}{\mathbf{d}}
\newcommand{\bde}{\mathbf{e}}
\newcommand{\bdf}{\mathbf{f}}
\newcommand{\bdg}{\mathbf{g}}
\newcommand{\bdh}{\mathbf{h}}

\newcommand{\bdp}{\mathbf{p}}
\newcommand{\bdq}{\mathbf{q}}
\newcommand{\bdr}{\mathbf{r}}
\newcommand{\bds}{\mathbf{s}}

\newcommand{\bdw}{\mathbf{w}}
\newcommand{\bdx}{\mathbf{x}}
\newcommand{\bdy}{\mathbf{y}}
\newcommand{\bdz}{\mathbf{z}}
\newcommand{\bdA}{\mathbf{A}}

\newcommand{\bdC}{\mathbf{C}}
\newcommand{\bdD}{\mathbf{D}}

\newcommand{\bdF}{\mathbf{F}}
\newcommand{\bdG}{\mathbf{G}}
\newcommand{\bdH}{\mathbf{H}}
\newcommand{\bdI}{\mathbf{I}}

\newcommand{\bdL}{\mathbf{L}}

\newcommand{\bdQ}{\mathbf{Q}}
\newcommand{\bdR}{\mathbf{R}}

\newcommand{\bdT}{\mathbf{T}}
\newcommand{\bdU}{\mathbf{U}}
\newcommand{\bdV}{\mathbf{V}}
\newcommand{\bdW}{\mathbf{W}}
\newcommand{\bdX}{\mathbf{X}}
\newcommand{\bdY}{\mathbf{Y}}
\newcommand{\bdZ}{\mathbf{Z}}

\newcommand{\bdgamma}{\boldsymbol{\gamma}}

\newcommand{\bdOmega}{\boldsymbol{\Omega}}
\newcommand{\bdomega}{\boldsymbol{\omega}}

\newcommand{\bdtheta}{\boldsymbol{\theta}}

\newcommand{\bdxi}{\boldsymbol{\xi}}
\newcommand{\bdSigma}{\boldsymbol{\Sigma}}
\newcommand{\bdGamma}{\boldsymbol{\Gamma}}

\newcommand{\bdLambda}{\boldsymbol{\Lambda}}

\newcommand{\stackeq}[1]{\stackrel{\text{(#1)}}{=}}

\newcommand{\stackgeq}[1]{\stackrel{\text{(#1)}}{\ge}}

\crefname{equation}{}{}
\crefname{figure}{Fig.}{Figs.}
\crefname{table}{Table}{Tables}
\crefname{myprop}{Proposition}{Propositions}
\crefname{mycorollary}{Corollary}{Corollarys}
\crefname{mylemm}{Lemma}{Lemmas}
\crefname{mytheorem}{Theorem}{Theorems}
\Crefname{secinapp}{Appendix}{Appendices}

\begin{document}

\title{Channel Estimation for LEO Satellite Massive MIMO OFDM Communications}

\author{
Ke-Xin~Li,~\IEEEmembership{Member,~IEEE,} 
~Xiqi~Gao,~\IEEEmembership{Fellow,~IEEE,} 
and~Xiang-Gen~Xia,~\IEEEmembership{Fellow,~IEEE} 

\thanks{Ke-Xin Li was with the National Mobile Communications Research Laboratory (NCRL), Southeast University, Nanjing 210096, and also with the Purple Mountain Laboratories (PML), Nanjing 211100, China. He is now with the Huawei Technologies Co., Ltd., Hangzhou 310053, China (e-mail: kkexinli@outlook.com).}
\thanks{Xiqi Gao is with the National Mobile Communications Research Laboratory (NCRL), Southeast University, Nanjing 210096, and also with the Purple Mountain Laboratories (PML), Nanjing 211100, China (e-mail: xqgao@seu.edu.cn).}
\thanks{Xiang-Gen Xia is with the Department of Electrical and Computer Engineering, University of Delaware,	Newark, DE 19716 USA (e-mail: xianggen@udel.edu).}
}

\maketitle

\begin{abstract}
	In this paper, we investigate the massive multiple-input multiple-output orthogonal frequency division multiplexing channel estimation for low-earth-orbit satellite communication systems.  First, we use the angle-delay domain channel to characterize the space-frequency domain channel. Then, we show that the asymptotic minimum mean square error (MMSE) of the channel estimation can be minimized if the array response vectors of the user terminals (UTs) that use the same pilot are orthogonal. Inspired by this, we design an efficient graph-based pilot allocation strategy to enhance the channel estimation performance.
	In addition, we devise a novel two-stage channel estimation (TSCE) approach, in which the received signals at the satellite are manipulated with per-subcarrier space domain processing followed by per-user frequency domain processing. Moreover, the space domain processing of each UT is shown to be identical for all the subcarriers, and an asymptotically optimal vector for the per-subcarrier space domain linear processing
	is derived. The frequency domain processing can be efficiently implemented by means of the fast Toeplitz system solver. 
	Simulation results show that the proposed TSCE approach can achieve a near performance to the MMSE estimation with much lower complexity.
\end{abstract}


\IEEEpeerreviewmaketitle

\section{Introduction}

The 5G and beyond 5G (B5G) networks are envisioned to support diverse usage scenarios, which have high requirement on transmission rate,  coverage area, reliability, latency, and access capability. Due to the expensive deployment costs, the current terrestrial 5G networks fail to provide ubiquitous services for the user terminals (UTs) in remote areas.  Thanks to the advantages of wide area coverage and immunity to natural disasters, satellite communications (SATCOM) have been a cornerstone to extend and complement terrestrial 5G networks \cite{Oltjon2021SatelliteSpace}. Nowadays, low-earth-orbit (LEO) satellites have been recognized as a promising infrastructure to provide high-throughput Internet access services to the UTs in unserved or underserved areas with shorter propagation delay, less pathloss, and lower launch costs, compared with the geostationary-earth-orbit (GEO) and medium-earth-orbit (MEO) competitors
\cite{Liu2021LEO5GBeyond,Oltjon2021SatelliteSpace,Fang20215GEmbraceSatellite}.

Multibeam satellites have received intensive attention in recent years \cite{Chrisopulos2015MultigroupFBSC}. 
By using a large number of spot beams generated at the satellite side, the frequency bands can be reused among beams with sufficient distance, which can significantly increase the system capacity for multibeam satellites.  
To further enhance the spectral efficiency, full frequency reuse (FFR) scheme, where all the spot beams use the same frequency bands, has been proposed for multibeam satellites \cite{Schwarz2019MIMOApplication}. Nevertheless, more sophisticated signal processing techniques, such as downlink (DL) precoding and uplink (UL) multi-user detection, are required to reduce the interference among adjacent beams \cite{Vazquez2016PrecodingChallenges}. 

Channel estimation plays an important role in SATCOM systems. For instance, in the UL data transmission phase, the satellite typically uses the multi-user detectors to recover the data symbols transmitted by multiple UTs so as to mitigate the multi-user interference. Whereas, the implementation of multi-user detectors depends critically on the channel state information (CSI) acquired by the satellite. Thus, to reap the performance gains brought by the multi-user detectors, the satellite needs to estimate the UTs' CSI with sufficient accuracy. 

In practical systems, CSI can be estimated through the training-based schemes, which relies on periodically inserted pilot signals known by the receiver \cite{You2016ChannelAcquisition}. 
By using the training-based schemes, the channel estimation and data detection can be decoupled, and the receiver's implementation complexity can be reduced \cite{Ma2005OptimalTrainingMIMO}. 
The channel estimation methods for the land mobile satellite (LMS) channels can be found in, for example, \cite{Arti2016HybridSatelliteTerrestrial,Arti2016ChannelEstimationDetectionLMS,Zhang2020UserActivity}, where it is assumed that the satellite is equipped with a single antenna. 
The channel estimation and pilot design for multibeam satellites were analyzed in  \cite{Michael2011ChannelEstimationMultibeam}, and the estimation performance was further enhanced in \cite{Wilfried2014NewResultsChannelEstimationMultibeam} by making use of the location information of UTs.

In conventional multibeam satellites, the beams generated by the satellites can only be modified at a very slow pace, which limits the adaptability to rapidly changing link conditions. In the past few years, massive multiple-input multiple-output (MIMO) systems have been made great success in terrestrial 5G networks \cite{Marzetta2010Noncooperative}. By deploying a large number of antennas at the base station (BS), massive MIMO can use multiple highly focused beams with flexible reconfigurability to serve tens of UTs simultaneously, thus improving the system spectral and energy efficiency significantly \cite{Hien2013ESEfficiency}. 
Nowadays, it has become possible to implement the fully digital beamforming at the satellite, which can generate agilely tunable beams \cite{Hong2017MultibeamAntenna5G}. In this paper,  we consider that the LEO satellite is equipped with a large number of antennas, and the beamforming therein is implemented in the fully digital domain, which can adapt to the dynamic link variations thanks to its flexible reconfigurability.  

Massive MIMO orthogonal frequency division multiplexing (OFDM) is an indispensable part of the future wideband wireless communication systems.
In recent years, OFDM is also considered in non-terrestrial networks (NTN) for 5G new radio (NR) \cite{3GPP_NonTerrestrial}.
A massive MIMO OFDM transmission approach for LEO SATCOM systems was proposed in \cite{You2019MassiveMIMOLEO}, which investigated the massive LEO satellite channel model, the DL/UL transmission schemes, and the user grouping strategies. 
The DL transmit design and UL transmit design for massive MIMO LEO SATCOM systems have been investigated in  \cite{KX2022DownlinkMassiveMIMOLEO} and \cite{KX2022UplinkMassiveMIMOLEO}, respectively.
In \cite{Roper2022DistributedDownlink}, the distributed downlink precoder and equalizer by using the low-dimensional angle information were proposed for multi-satellite systems, which can reduce the inter-satellite coordination overhead.

The channel estimation has been extensively studied for small-scale MIMO OFDM \cite{Li2002SimplifiedChannelEstimation,Barhumi2003OptimalTraining,Minn2006OptimalTraining} and massive MIMO OFDM \cite{You2015PilotReuse,You2016ChannelAcquisition,Sheng2017OptimalTraining,Liu2018DownlinkHiddenMarkovian,Kuai2019StructuredTurbo,Zhang2021UnifyingMessage,Liu2021SparseChannel} communications in terrestrial wireless systems.
However, the exiting techniques cannot be straightforwardly adopted  for massive MIMO OFDM LEO satellite channel estimation due to the following reasons. Firstly, the Doppler shift and propagation delay of LEO satellite channel are much higher than those of the terrestrial wireless channel, which makes it difficult for reliable channel estimation. Secondly, the existing channel estimation techniques require the received signals to be jointly processed in the space-frequency domain, which can bring considerable computational overhead for capability-limited LEO satellite payloads. Recently, a deep learning based channel prediction is devised in \cite{Zhang2021DeepLearingLEO} to address the channel outdating effects. In addition, the orthogonal time frequency space (OTFS) modulation is resorted to combat the high Doppler shift in LEO satellite communications, e.g., see \cite{Shen2022RandomAccessOTFSLEO,Wang2022JointCEDTOTFSLEO}, and  space-air-ground integrated networks (SAGIN) \cite{Xu2022OTFSAidedRIS}. So far, the channel estimation for massive MIMO OFDM LEO SATCOM systems has not been investigated specifically.

In this paper, we investigate the channel estimation for massive MIMO OFDM LEO SATCOM by making full use of the LEO satellite channel properties, in which the satellite is equipped with uniform planar array (UPA) and each UT has a single antenna. The relatively large Doppler shift and propagation delay are pre-compensated at each UT to support the massive MIMO OFDM transmission.
Our major contributions are summarized as follows.
\begin{enumerate}
		\item Starting from the signal model after the Doppler and delay compensations, together with the physical channel models, we analyze the channel characteristics over one OFDM symbol for LEO satellite massive MIMO OFDM communications. Specifically, we derive the angle-delay domain channel  (ADC) to represent the space-frequency domain channel (SFC). The ADC allows fine sampling in the delay domain to achieve an accurate approximation of the SFC, and paves the way for better solving the channel estimation problem. 
		\item With the ADC based SFC representation, we investigate the minimum mean square error (MMSE) channel estimation with pilot reuse. We obtain the conditions for asymptotic MMSE	minimization, which indicates under what conditions the pilot reuse is asymptotically optimal. Specifically, we prove that the asymptotic MMSE can be minimized if the array response vectors of the UTs that use the same pilot are orthogonal with each other. Inspired by this, we devise an efficient graph-based pilot allocation strategy, so that the channel estimation performance can be enhanced under pilot reuse.
		\item We prove that the MMSE estimator has a two-stage structure in the low SNR and high SNR regimes. Motivated by this, we propose a novel two-stage channel estimation (TSCE) approach for the general case, in which the channel estimation is fulfilled with per-subcarrier space domain processing followed by per-user frequency domain processing. Moreover, the space domain processing of each UT is identical for all the subcarriers, and an asymptotically optimal vector for the per-subcarrier space domain linear processing is derived. By utilizing the fast Toeplitz system solver, the per-user frequency domain processing can be efficiently implemented. Simulation results show that the TSCE approach can achieve a near optimal performance with much lower computational complexity.
\end{enumerate}

The remainder of this paper is organized as follows. \Cref{section_system_model} introduces the system model, where the ADC is obtained to characterize the SFC. In \Cref{section_channel_estimation}, the pilot reuse for channel estimation is investigated, and an efficient pilot allocation strategy is devised. In \Cref{section_two_stage_channel_estimation}, the TSCE approach is proposed, which can achieve a near optimal performance with much lower computational complexity. \Cref{section_simulation_results} verifies the proposed approach with the simulation results, and \Cref{section_conclusion} concludes this paper.

\textit{Notations:} Throughout this paper, lower case letters denote scalars, and boldface lower (upper) case letters denote vectors (matrices). The set of all $n$-by-$m$ complex (real) matrices is denoted as $\bbC^{n\times m}$ ($\bbR^{n\times m}$). The trace, vectorization, inverse, Moore-Penrose inverse, conjugate, transpose, and conjugate transpose for matrix are represented by $\trace(\cdot)$,  $\Xvec{\cdot}$, $(\cdot)^{-1}$, $(\cdot)^{\dagger}$, $(\cdot)^*$, $(\cdot)^T$, and $(\cdot)^H$, respectively. The Euclidean norm of vector $\bdx$ is denoted as $\snorm{\bdx} = \sqrt{\bdx^H \bdx}$. The Frobenius norm of matrix $\bdX$ is denoted by $\snorm{\bdX}_{\mathrm{F}} = \sqrt{\trace(\bdX\bdX^H)}$. $\sabs{\clA}$ denotes the cardinality of the set $\clA$. The identity matrix is represented by $\bdI$ or $\bdI_N$. $\bdF_N$ denotes the $N$-dimensional discrete Fourier transform (DFT) matrix. $\otimes$ denotes the Kronecker product. $[\bdX]_{p,q}$ represents the $(p,q)$th element of matrix $\bdX$. The diagonal matrix with $\bdx$ along its main diagonal is denoted as $\diag\{\bdx\}$. $\lceil x \rceil$ and $\lfloor x \rfloor$ denote rounding $x$ to the nearest integers towards $\infty$ and $-\infty$, respectively. $\modd(x,N)$ denotes $x$ modulo $N$. $[\bdx_i]^{\rR}_{i\in\clI}$ and $[\bdx_i]^{\rC}_{i\in\clI}$ denote arranging $\bdx_i$'s along rows and columns, respectively, according to the increasing order of the indices in $\clI$. $\bbE \{ \cdot \}$ denotes the mathematical expectation. $\clCN(\bdzro,\bdSigma)$ represents the circularly symmetric complex Gaussian distribution with zero mean and covariance matrix $\bdSigma$.

\section{System  Model} \label{section_system_model}
In this section, we first describe the configuration of LEO satellite massive MIMO OFDM system. Then, by using the Doppler and delay compensation techniques, we derive the signal model for OFDM transmission. After that, we build the LEO satellite channel model that represents the SFC with ADC, and analyze the channel's statistical properties.

\subsection{System Configuration}

\begin{figure}[!t]
	\centering
	\includegraphics[width=0.48\textwidth]{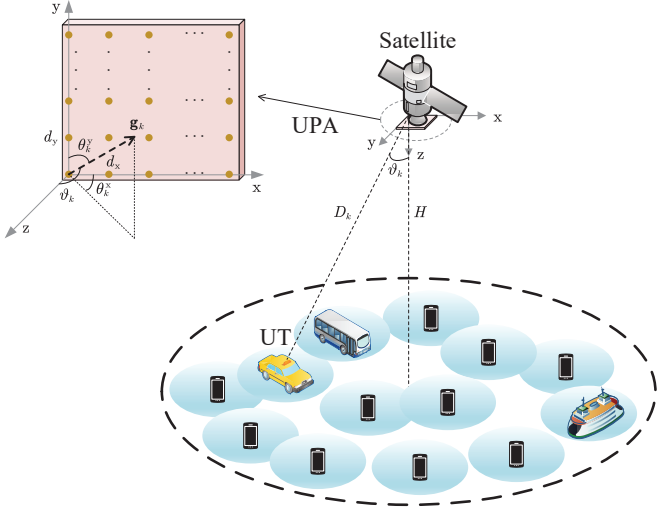}
	\caption{LEO satellite massive MIMO OFDM system.}
	\label{fig_MIMO_OFDM_Satellite}
\end{figure}

We consider an FFR wideband massive MIMO LEO satellite system operating over lower frequency bands, e.g., L/S/C bands. 
As shown in \Cref{fig_MIMO_OFDM_Satellite}, the UPA at the satellite has $\Mx$ and $\My$ directional antenna elements in the $\rx$-axis and $\ry$-axis, respectively. The total number of antenna elements at the satellite is $\Mx \My \triangleq M$.
The LEO satellite serves $K$ mobile UTs, and each UT has a single omnidirectional antenna. The set of UT indices is denoted by $\clK = \{ 1,2,\dots,K \}$.

The OFDM modulation is used to support wideband transmission in LEO SATCOM systems, where the frequency selective fading channel is transformed into a collection of parallel flat fading channels. The total number of subcarriers is $\Nc$, and  the subcarrier spacing is $\Delta f$. The system sampling interval is denoted by $\Ts = 1/(\Nc \Delta f)$. The guard interval, a.k.a., circular prefix (CP), with a length of $\Ng$ is appended at the beginning of each OFDM symbol. The time interval of CP is given by $\Tg = \Ng \Ts$. 
The time intervals of one OFDM symbol without and with CP are denoted by $\Tc = \Nc \Ts$ and $T = \Tg+\Tc$, respectively.

\begin{figure}[!t]
	\centering
	\includegraphics[width=0.42\textwidth]{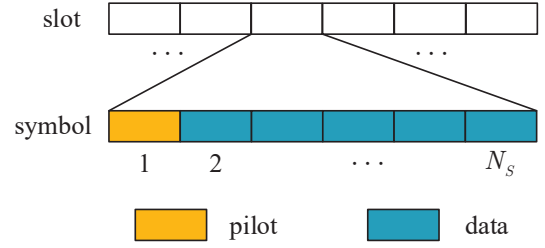}
	\caption{UL frame structure.}
	\label{fig_frame_structure}
\end{figure}

The massive MIMO LEO SATCOM system is assumed to operate in the frequency division duplexing (FDD) mode, where the UL and DL occupy different frequency bands.  The UL frame structure is shown in \Cref{fig_frame_structure}. Specifically, the time resource in the UL phase is divided into a number of slots, and each slot consists of $N_S$ OFDM symbols \cite{3GPP_NR_PhysicalChannels}. Moreover, the first OFDM symbol in each slot is used for the UL channel estimation, while the remaining $N_S-1$ OFDM symbols are used for the UL data transmission.

For the training-based UL channel estimation strategies, the mobile UTs on the ground transmit pilot sequences to a single LEO satellite simultaneously. 
The pilot sequences are perfectly known by the satellite and the UTs. After receiving the pilot signals, the satellite aims to estimate the channel parameters of all the UTs. 

\subsection{Signal Models}\label{subsec_signal_model}
The received signal $\bdy(t) \in \Complex{M}{1}$ at the satellite at time instant $t$ is given by
\begin{equation}\label{signal_model_analog_baseband}
	\bdy(t) = \sum_{k=1}^K \int_{-\infty}^{\infty} \ckbdh_k(t,\tau) x_k(t-\tau) \dint \tau + \bdz(t)\comma 
\end{equation}
where $\ckbdh_k(t,\tau) \in \Complex{M}{1}$ and $x_k(t) \in \bbC$ are the time-varying channel impulse response and transmit signal of UT $k$, respectively, and $\bdz(t) \in \Complex{M}{1}$ is the additive noise signal encountered by the satellite. Moreover, the channel impulse response $\ckbdh_k(t,\tau)$ can be expressed as
\begin{equation}\label{channel_model_analog_baseband}
	\ckbdh_k(t,\tau) = \sum_{q=1}^{Q_k} \cka_{k,q} e^{j2\pi \nu_{k,q} t} \delta(\tau - \tau_{k,q}) \cdot \bdg_{k,q}\comma
\end{equation}
where  $\delta(x)$ is the Dirac delta function, $Q_k$ is the number of paths of UT $k$'s channel, $\cka_{k,q}$, $\nu_{k,q}$,  $\tau_{k,q}$ and $\bdg_{k,q}$ are the complex channel gain, Doppler shift, propagation delay and array response vector at the satellite side, associated with the $q$th path of UT $k$'s channel, respectively. 

In LEO SATCOM systems, the Doppler shifts and propagation delays are much more serious than those in terrestrial wireless networks, because of the large relative moving velocity and high altitude of the satellite, which can bring considerable challenges on frequency and time synchronization. Therefore, the characteristics of the Doppler shifts and propagation delays in LEO SATCOM systems need to be well understood, so that the compensation techniques can be used to support the wideband transmission. 
The Doppler shift $\nu_{k,q}$ for the $q$th path of UT $k$'s channel is composed of two terms as $\nu_{k,q} = \nu_{k,q}^{\sat} + \nu_{k,q}^{\ut}$ \cite{Papath2001Acomparison}, where $\nu_{k,q}^{\sat}$ and $\nu_{k,q}^{\ut}$ are the Doppler shifts induced by the movement of the satellite and UT $k$, respectively. Furthermore, $\nu_{k,q}^{\sat}$ is approximately identical for all the paths of UT $k$'s channel \cite{Papath2001Acomparison}, i.e., $\nu_{k,q}^{\sat} = \nu_k^{\sat}$, $q=1,\dots,Q_k$. In addition, the propagation delay of UT $k$ can also be written as $\tau_{k,q} = \tau_{k}^{\sat} + \tau_{k,q}^{\ut}$, where $\tau_{k}^{\sat}$ is the large propagation delay part due to the long distance between the satellite and UT $k$, and $\tau_{k,q}^{\ut}$ denotes the residual propagation delay part depending on the scattering environment around UT $k$.

Let $\bdtheta_{k,q}=(\theta_{k,q}^{\rx},\theta_{k,q}^{\ry})$ denote the paired angles-of-arrival (AoAs) for the $q$th path of UT $k$'s channel. 
Then, the array response vector $\bdg_{k,q}$ can be written as $\bdg_{k,q} = \bdg(\bdtheta_{k,q})$. For arbitrary $\bdtheta = (\theta_{\rx},\theta_{\ry})$, the vector $\bdg(\bdtheta)$ is defined as $\bdg(\bdtheta) = \bda_{\Mx} ( \sin \theta_{\ry} \cos \theta_{\rx} ) \otimes 
\bda_{\My} ( \cos \theta_{\ry} )$.
Here, $\bda_{n_{\rv}} ( x ) \in \Complex{n_{\rv}}{1}$ is given by $\bda_{n_{\rv}} \left( x \right) = \frac{1}{\sqrt{n_{\rv}}} [1\ e^{-j \frac{2\pi d_{\rv}}{\lambda} x }\ \cdots \ e^{-j\frac{2\pi d_{\rv}}{\lambda} (n_{\rv}-1) x } ]^T$,
where $j = \sqrt{-1}$, $\lambda=c/f_c$ is the wavelength, $c$ is the speed of light, $f_c$ is the carrier frequency, $d_{\rv}$ is the spacing between adjacent antennas along the $\rv$-axis with  $\rv \in \{ \rx,\ry\}$. 
As a result of the long distance between the satellite and UT $k$, the paired AoAs $\bdtheta_{k,q}$ for different paths of UT $k$'s channel tend to be identical, i.e., $\bdtheta_{k,q} = \bdtheta_k$, $q = 1,\dots,Q_k$. Thus, we can discard the subscript of path $q$ in $\bdg_{k,q}$ and rewrite it as $\bdg_{k,q} = \bdg_{k} = \bdg(\bdtheta_k)$, $q = 1,\dots,Q_k$, where $\bdtheta_k = (\theta_k^{\rx},\theta_k^{\ry})$ is referred to as the paired AoAs of UT $k$. For convenience, let us denote the paired space angle of UT $k$ as $\bdxi_k = (\xi_k^{\rx},\xi_k^{\ry})$, where $\xi_k^{\rx} = \sin \theta_k^{\ry} \cos \theta_k^{\rx} $ and $\xi_k^{\ry} = \cos \theta_k^{\ry}$.
The nadir angle of UT $k$ is defined as $\vtheta_k = \cos^{-1}( \sin \theta_k^{\ry} \sin \theta_k^{\rx} )$, and its maximum value is denoted by $\vtheta_{\max}$. Moreover, the nadir angle $\vtheta_k$ and paired space angle $\bdxi_k$ satisfy the relation $
		\cos \vtheta_k  = \sin \theta_k^{\ry} \sin \theta_k^{\rx} = \sqrt{1 - (\sin \theta_k^\ry \cos \theta_k^{\rx})^2 - (\cos\theta_k^{\ry})^2}
		= \sqrt{1 - (\xi_k^{\ry})^2 - (\xi_k^{\rx})^2} \ge \cos \vtheta_{\max}$.
Hence, the paired space angle $\bdxi_k$ of UT $k$ should be located in the circle region $\{(x,y)| x^2 + y^2 \le \sin^2\vtheta_{\max}\}$. Owing to the long distance between the LEO satellite and UTs, the space angle pairs $\{\bdxi_k\}_{k=1}^K$ change quite slowly. Therefore, it is reasonable to assume that the space angle pairs can be perfectly known by the satellite and UTs.

Let $\{x_{k,s,r}\}_{r=0}^{\Nc-1}$ denote the frequency domain transmit signal of UT $k$ in the $s$th OFDM symbol. The time domain transmit signal of UT $k$ is given by 
\begin{align}
	x_{k,s}(t) = \sum_{r=0}^{\Nc-1} x_{k,s,r} e^{j2\pi r \Delta f \cdot t }\comma \ -\Tg \le t-sT \le \Tc.
\end{align} 
Let us denote $\nu_{k}^{\cps} = \nu_{k}^{\sat}$ and $\tau_{k}^{\cps} = \tau_{k}^{\sat}$, respectively. The Doppler shift $\nu_k^{\sat}$ and the propagation delay $\tau_k^{\sat}$ are dependent on the locations of the satellite and UT $k$. The UTs can acquire their location information, e.g., by using the global positioning system (GPS), and then $\nu_k^{\sat}$'s and $\tau_k^{\sat}$'s can be directly computed with the help of the ephemeris.
By using the Doppler shift and delay compensation techniques, the time domain transmit signal of UT $k$ after compensation is given by $x_{k,s}^{\cps}(t) = x_{k,s}(t+\tau_{k}^{\cps})e^{-j2\pi \nu_{k}^{\cps}(t+\tau_{k}^{\cps})}$.
The time domain received signal at the satellite in the $s$th OFDM symbol is given by
\begin{align} \label{ys_cps}
	\bdy_{s}^{\cps}(t) &= \sum_{k=1}^K \int_{-\infty}^{\infty} \ckbdh_k(t,\tau) x_{k,s}^{\cps}(t-\tau) \dint \tau + \bdz_{s}(t)\notag \\
	&= \sum_{k=1}^K \sum_{q=1}^{Q_k} \cka_{k,q} e^{j2\pi \nu_{k,q} t} x_{k,s}(t -\tau_{k,q}^{\ut}) e^{-j2\pi \nu_{k}^{\cps}(t-\tau_{k,q}^{\ut})} \notag \\
	&\qquad \qquad\cdot \bdg_{k}  + \bdz_{s}(t) \notag \\
	&= \sum_{k=1}^K \sum_{q=1}^{Q_k}\bva_{k,q} e^{j2\pi \nu_{k,q}^{\ut} t} x_{k,s}(t -\tau_{k,q}^{\ut})\cdot \bdg_{k}  + \bdz_{s}(t) \notag \\
	&= \sum_{k=1}^K \int_{-\infty}^{\infty} \bvbdh_k(t,\tau) x_{k,s}(t-\tau) \dint \tau + \bdz_{s}(t) \comma
\end{align}
where $\bdz_s(t)$ is the additive noise. Here, $\bvbdh_k(t,\tau)$ is the effective channel impulse response of UT $k$, and it is given by
\begin{align}
	\bvbdh_k(t,\tau) = \sum_{q=1}^{Q_k} \bva_{k,q} e^{j2\pi \nu_{k,q}^{\ut} t} \delta(\tau - \tau_{k,q}^{\ut}) \cdot \bdg_{k}\comma
\end{align}
where $\bva_{k,q}$ is the effective channel gain for the $q$th path of UT $k$'s channel. In OFDM systems, to avoid the inter-symbol interference, $\Ng$ should be properly chosen to satisfy $\Ng > \tau_{k,q}^{\ut}/\Ts$, $\forall k,q$. It is worth noting that compared with the original channel impulse response $\ckbdh_k(t,\tau)$, the influences of Doppler shifts and propagation delays in the effective channel impulse response $\bvbdh_k(t,\tau)$ have been significantly alleviated. By doing so, the variation rate and propagation delays of the effective channels can be much lower than those of the original ones, which means that the effective channels can be treated as block fading and synchronized in time and frequency hereafter.

Moreover, let us denote $\bdh_k(t,f)$ as the effective channel frequency response of UT $k$, and it can be derived by $\bdh_k(t,f) = \int_{-\infty}^{\infty} \bvbdh_k(t,\tau) e^{-j 2 \pi f \tau } \dint \tau 
= d_{k}(t,f) \bdg_k$,
where $d_k(t,f)$ is defined as $d_k(t,f) = \sum_{q=1}^{Q_k} \bva_{k,q} e^{j2\pi (\nu_{k,q}^{\ut} t - f\tau_{k,q}^{\ut})}$.
The frequency domain received signal at the satellite over the $r$th subcarrier of the $s$th OFDM symbol can be written as
\begin{align}\label{ysr_sumk_hksr_xksr_zsr}
	\bdy_{s,r} &= \xinv{\Tc} \int_{sT}^{sT+\Tc} \bdy_{s}^{\cps}(t) e^{-j2\pi r \Delta f \cdot t} \dint t  \notag \\
	&= \sum_{k=1}^K \bdh_{k,s,r} x_{k,s,r} + \bdz_{s,r}\comma
\end{align}
where $\bdh_{k,s,r}$ and $\bdz_{s,r}$ are the channel vector of UT $k$ and the additive noise, respectively, both over the $r$th subcarrier of the $s$th OFDM symbol. 
The channel vector $\bdh_{k,s,r}$ in \Cref{ysr_sumk_hksr_xksr_zsr} can be further written as
\begin{equation} \label{channel_freq_resp_UT_k_akq}
	\bdh_{k,s,r} = \bdh_{k}(sT,r\Delta f) =  d_{k,s,r} \bdg_k\comma
\end{equation}
where $d_{k,s,r} = d_k(sT,r\Delta f)$.

Notice that in terrestrial wireless communications, the angular spread of each UT's channel observed by the BS is usually nonzero and depends on the scatter distribution in practical propagation environment \cite{Liu2018DownlinkHiddenMarkovian,Liu2021SparseChannel,You2016ChannelAcquisition}. However, in LEO SATCOM systems, the angular spread of each UT's channel at the satellite side is zero \cite{3GPP_NonTerrestrial,You2019MassiveMIMOLEO,KX2022DownlinkMassiveMIMOLEO}. Hence, the space domain characteristics of LEO satellite channels are relatively deterministic, which makes it possible to develop effective low-complexity channel estimation techniques.

\subsection{Channel Model}\label{subsec_channel_model}
In this subsection, we analyze the channel characteristics over one OFDM symbol occupied by the pilot. For brevity, the subscript of OFDM symbol $s$ is omitted afterwards. 
The channel vector of UT $k$ on the $r$th subcarrier of a specific OFDM symbol can be written as $\bdh_{k,r} = d_{k,r}\bdg_k$,
where $d_{k,r} = \sum_{q=1}^{Q_k} a_{k,q} e^{-j2\pi r \Delta f \cdot \tau_{k,q}^{\ut}}$, and $a_{k,q}$ is the effective channel gain for the $q$th path of UT $k$'s channel within the considered OFDM symbol.
It is assumed that $\Np$ subcarriers are used for the channel estimation, with the corresponding index set given by $\clN_{\rp} \triangleq \{r_{\rp},\dots,r_{\rp}+\Np-1\} \subseteq  \clN_{\rc} \triangleq \{ 0,\dots,\Nc-1 \}$. 
Let us denote $\bdH_{\rp,k} = [ \bdh_{k,r_{\rp}}\ \cdots\ \bdh_{k,r_{\rp}+\Np-1} ] \in \Complex{M}{\Np}$ as the channel matrix of UT $k$ over these $\Np$ subcarriers, and $\bdH_{\rp,k}$ can be written as
\begin{align} \label{SFCM}
	\bdH_{\rp,k} = \bdg_k \bdd_{\rp,k}^T\comma
\end{align}
where $\bdd_{\rp,k} = [ d_{k,r_{\rp}} \ \cdots \ d_{k,r_{\rp}+\Np-1} ]^T \in \Complex{\Np}{1}$. Hereafter, $\bdH_{\rp,k}$ is referred to as the SFC of UT $k$. Unlike the SFC in terrestrial wireless communications \cite{You2016ChannelAcquisition,Liu2021SparseChannel}, the SFC in LEO SATCOM systems exhibits the rank-one structure.
In addition, $\bdd_{\rp,k}$ can be further written as
\begin{align}
	\bdd_{\rp,k} = \sum_{q=1}^{Q_k} a_{k,q} \bdp(\tau_{k,q}^{\ut})\comma
\end{align}
where $\bdp(x) = [ e^{-j2\pi r_\rp \Delta f \cdot x  } \ \cdots \ e^{-j2\pi (r_\rp+\Np-1) \Delta f \cdot x } ]^T \in \Complex{\Np}{1}$.
By noticing the fact that $\tau_{k,q}^{\ut} \in [0,\Tg)$, $\forall k ,q$, the CP time duration $[0,\Tg)$ is partitioned into $\Nd$ intervals as follows
\begin{equation} \label{Tg_Nd_interval}
	[0,\Tg) = \clD_0 \cup \cdots \cup \clD_{\Nd-1} \comma
\end{equation}
where $\clD_{\ell} = \left[\tau_{\ell}, \tau_{\ell+1}\right)$, $\tau_{\ell}$ is the $\ell$th grid point given by $\tau_{\ell} = \ell \Ld /(\Nd \Np \Delta f)$ with $\Ld = \xceil{\Np\Ng/\Nc}$, $\ell \in\{0, \dots, \Nd-1\} \triangleq \clN_{\rd}$, and $\tau_{\Nd} = \Tg$. For convenience, let us define $\mud = \Nd/\Ld$ as the refining factor.
The physical meaning of $\Nd$ will be clear later.
By using the $\Nd$ intervals in \eqref{Tg_Nd_interval}, $\bdd_{\rp,k}$ can be rewritten as
\begin{align}
	\bdd_{\rp,k} = \sum_{\ell=0}^{\Nd-1} \sum_{q\in\clQ_{k,\ell}} a_{k,q} \bdp(\tau_{k,q}^{\ut})\comma
\end{align}
where $\clQ_{k,\ell} = \{q| \tau_{k,q}^{\ut} \in \clD_{\ell}\}$.
For the cases that $\Nd$ is sufficiently large, $\bdp(\tau_{k,q}^{\ut})$ with $\tau_{k,q}^{\ut} \in \clD_{\ell}$ can be well approximated by $\bdp(\tau_{\ell})$, and consequently $\bdd_{\rp,k}$ can be well approximated by
\begin{align} \label{dpk_dtk}
	\bdd_{\rp,k} = \sum_{\ell=0}^{\Nd-1} \alpha_{k,\ell} \bdp(\tau_{\ell}) 
	= \FNpNd \bdd_{\rt,k}\comma
\end{align}
where $\FNpNd = [\bdp(\tau_0) \ \cdots \ \bdp(\tau_{\Nd-1})] \in \Complex{\Np}{\Nd}$ and $\bdd_{\rt,k} = [ \alpha_{k,0} \ \cdots \ \alpha_{k,\Nd-1} ]^T \in \Complex{\Nd}{1}$ with $\alpha_{k,\ell} = \sum_{q\in\clQ_{k,\ell}} a_{k,q}$. 
Notice that $\Nd$ determines how accurately $\bdp(\tau_{k,q}^{\ut})$ is approximated by $\bdp(\tau_{\ell})$. The larger $\Nd$ is, the more accurate the approximation is.
Therefore, the SFC $\bdH_{\rp,k}$ in \eqref{SFCM} can be further written as
\begin{align}
	\bdH_{\rp,k} = \bdg_k \bdd_{\rt,k}^T \FNpNd^T.
\end{align}
Henceforth, $\bdd_{\rt,k}$ is referred to as the ADC of UT $k$. 

Throughout this paper, it is assumed that $\alpha_{k,\ell}$ can be modeled as $\alpha_{k,\ell} = \sqrt{\beta_k} \eta_{k,\ell}$,
where $\beta_k$ is the large-scale fading parameter of UT $k$, and $\eta_{k,\ell}$ accounts for the fast small-scale fading parameter for the $\ell$th tap of UT $k$'s channel. In addition, $\eta_{k,\ell}$ is assumed to be distributed as $\eta_{k,\ell} \sim \mathcal{CN} ( 0, \gamma_{k,\ell})$ with $\sum_{\ell=0}^{\Nd-1} \gamma_{k,\ell} = 1$, and $\{\gamma_{k,\ell}|\forall \ell\}$ is referred to as the power delay profile (PDP) of UT $k$. Moreover, we assume that $\alpha_{k,\ell}$'s are independent for different $k$ and $\ell$, i.e., $\bbE\{ \alpha_{k,\ell} \alpha_{k',\ell'}^* \} = \beta_k \gamma_{k,\ell} \delta(k-k') \delta(\ell-\ell')$.
Let us define the angle-delay domain channel correlation matrix $\bdR_{\rt,k} \in \Real{\Nd}{\Nd}$ as $\bdR_{\rt,k} = \bbE \{ \bdd_{\rt,k} \bdd_{\rt,k}^H  \} = \diag \{ \bdomega_k \}$,
where $\bdomega_k = [\omega_{k,0} \ \cdots \ \omega_{k,\Nd-1}]^T \in \Real{\Nd}{1}$ with $\omega_{k,\ell} = \beta_k \gamma_{k,\ell}$. 
In this paper, it is assumed that the slow-varying channel parameters $\{(\bdxi_k, \bdomega_{k})|\forall k\}$ are perfectly known by the satellite and UTs. 

\section{Pilot Reuse for Channel Estimation} \label{section_channel_estimation}
In this section, we study the pilot reuse for channel estimation in LEO satellite massive MIMO OFDM systems. 
First, building on the ADC in \Cref{subsec_channel_model}, we derive the MMSE channel estimation with pilot reuse. Then, we obtain the conditions for asymptotic MMSE minimization, which indicates under what conditions the pilot reuse is asymptotically optimal. Inspired by the conditions, we devise a graph-based pilot allocation algorithm, which allows efficient implementation and can achieve near-optimal performance.

\subsection{Pilot Reuse} \label{subsec_pilot_reuse}

In practice, the number of available pilots is much less than that of UTs in the coverage area of an LEO satellite. Therefore, pilot reuse would be an indispensable part of LEO SATCOM systems.
Intuitively, the same pilot can be used by the UTs that would not interfere seriously with each other. In this subsection, we will formally state the pilot reuse for LEO satellite massive MIMO OFDM communications.

Let $S$ denote the number of pilots, and the set of pilots is given by $\clX_{\rp} = \{\bdX_{\rp,1},\dots,\bdX_{\rp,S}\}$. 
The set of indices of UTs that use the pilot signal $\bdX_{\rp, s}$ is denoted by $\clK_{s}$ with $K_s = \sabs{\clK_s}$, where $\cup_{s=1}^S \clK_s = \clK$ and $\clK_{s} \cap \clK_{s'} = \emptyset$, $\forall s \ne s'$.
Let us denote $\bdY_{\rp} \in \Complex{M}{\Np}$ as the UL received signal at the satellite over the $\Np$ subcarriers, and it can be written as
\begin{equation}
	\bdY_{\rp} = \sum_{k=1}^{K} \sqrt{\frac{P}{\Np}} \bdH_{\rp,k} \bdX_{\rp, s_k} + \bdZ_{\rp}\comma \label{Y_RxSignal_allSubcarrier}
\end{equation}
where $P$ is the transmit power of each UT, $s_k$ is the index of pilot used by UT $k$, and $\bdZ_{\rp} \in \Complex{M}{\Np}$ is the additive complex Gaussian noise whose elements are independent and identically distributed as $\clCN(0,\sigma^2)$. 

By following the spirits of phase shift pilots \cite{You2016ChannelAcquisition}, the $s$th pilot $\bdX_{\rp, s}$ can be written as
\begin{align}
	\bdX_{\rp,s} = \diag \left\{ e^{-j\frac{2\pi \phi_s r_{\rp}}{\Npe} }, \dots, e^{-j\frac{2\pi \phi_s (r_{\rp}+\Np-1)}{\Npe} } \right\} \bdX_{\rc} \comma \label{Xps_phase_shift_pilot}
\end{align}
where $\Npe = \Np \mud$, $\phi_s \in \{0,\dots,\lfloor\Npe\rfloor-1\}$ satisfies $\modd (\phi_{s'} - \phi_{s}, \lfloor\Npe\rfloor) \ge \Nd$, $\forall s \ne s'$. In addition, $\bdX_{\rc} = \diag\{\bdx_{\rc}\}$ is the basic pilot with unit modulus elements, i.e., $\bdX_{\rc} \bdX_{\rc}^H = \bdI$. For example, $\bdX_{\rc}$ can be constructed based on the Zadoff-Chu sequences, which have been extensively used in 5G NR \cite{Dahlman20185GNR}. Moreover, $S$ should be limited by $S \le \lfloor \lfloor\Npe\rfloor/\Nd \rfloor$.

Because the paired space angles $\{\bdxi_k|\forall k\}$ as well as the array response vectors $\{\bdg_k|\forall k\}$ are known by the satellite, the estimation of the SFC $\{\bdH_{\rp,k}|\forall k\}$ can be converted into that of the vectors $\{\bdd_{\rp,k}|\forall k\}$. In addition, owing to the fact that the ADC can better characterize the SFC with fewer parameters, we first consider estimating the ADC $\{\bdd_{\rt,k}|\forall k\}$ in the following, which is one of the key differences between terrestrial and satellite massive MIMO systems.

Let $\bdy_{\rp} \triangleq \Xvec{\bdY_{\rp}\bdX_{\rc}^*} \in \Complex{M\Np}{1}$ and $\bdz_{\rp} \triangleq \Xvec{\bdZ_{\rp}\bdX_{\rc}^*} \in \Complex{M\Np}{1}$. From \eqref{Y_RxSignal_allSubcarrier}, it can be derived that
\begin{align}
	\bdy_{\rp} &= \sum_{k=1}^{K} \sqrt{\frac{P}{\Np}} \Xvec{ \bdg_{k} \bdd_{\rp,k}^T \bdX_{\rp,s_k} \bdX_{\rc}^* } + \bdz_{\rp} \notag \\
	&\stackeq{a} \sum_{k=1}^{K} \sqrt{\frac{P}{\Np}} \left( \bdX_{\rp,s_k} \bdX_{\rc}^* \FNpNd \otimes \bdg_k \right) \bdd_{\rt,k} + \bdz_{\rp} \notag \\		
	&\stackeq{b} \sum_{k=1}^{K} \sqrt{\frac{P}{\Np}} \left( \bdF_{\rp,\re} \bdD_{\rd,s_k} \otimes \bdg_k \right) \bdd_{\rt,k} + \bdz_{\rp} \notag \\
	&= \bdA_{\rp} \bdd_{\rt} + \bdz_{\rp}\comma \label{signal_model_kronecker}
\end{align}
where (a) comes from the relation in \eqref{dpk_dtk}, (b) follows from the fact that $\bdX_{\rp,s_k} \bdX_{\rc}^* \FNpNd$ can be rewritten as 
\begin{align}
	&\bdX_{\rp,s_k} \bdX_{\rc}^* \FNpNd \notag \\
	={}& \underbrace{\left[\begin{matrix}
			\bdzro_{\Np\times r_\rp} & \bdI_{\Np} & \bdzro_{\Np\times R_{\rp}}
		\end{matrix}\right]}_{\bdD_{\rp}} \bdF_{\Npe} \underbrace{\left[\begin{matrix}
			\bdzro_{\phi_{s_k}\times\Nd} \\
			\bdI_{\Nd} \\
			\bdzro_{\Phi_{s_k}\times\Nd}
		\end{matrix}\right]}_{\bdD_{\rd,s_k}} \notag \\
	={}& \bdF_{\rp,\re} \bdD_{\rd,s_k}\comma
\end{align}
with $R_{\rp} = \Npe-r_\rp-\Np$, $\Phi_{s_k} = \Npe-\phi_{s_k}-\Nd$, $\bdF_{\rp,\re} = \bdD_{\rp} \bdF_{\Npe} \in \Complex{\Np}{\Npe}$, $\bdA_{\rp} \in \Complex{M\Np}{K\Nd}$ is defined as $\bdA_{\rp} = \sqrt{\frac{P}{\Np}} \left[\bdF_{\rp,\re} \bdD_{\rd,s_1} \otimes \bdg_1 \ \cdots \ \bdF_{\rp,\re} \bdD_{\rd,s_K} \otimes \bdg_K\right]$,
and $\bdd_{\rt} = [\bdd_{\rt,1}^T \ \cdots \ \bdd_{\rt,K}^T]^T \in \Complex{K\Nd}{1}$.
In addition, $\bdz_{\rp}$ is distributed as $\bdz_{\rp} \sim \clCN(\bdzro,\sigma^2 \bdI)$.

Then, once the estimate of the vector $\bdd_{\rt}$ is obtained, the estimate of the vector $\bdd_{\rp} = [\bdd_{\rp,1}^T \ \cdots \ \bdd_{\rp,K}^T]^T \in \Complex{K\Nd}{1}$ can be immediately derived via
\begin{align}
	\htbdd_{\rp} = (\bdI_K \otimes \FNpNd ) \htbdd_{\rt}.
\end{align}

According to \eqref{signal_model_kronecker}, the general MMSE estimation of $\bdd_{\rt}$ is given by \cite{Kailath2000LinearEstimation}
\begin{align}
	\htbdd_{\rt} =\Xinv{\bdR_{\rt} \bdA_\rp^H \bdA_\rp + \sigma^2 \bdI} \bdR_{\rt} \bdA_\rp^H \bdy_{\rp} \comma \label{dfk_estimate_MMSE}
\end{align}
where $\bdR_{\rt} = \diag\left\{[\bdomega_1^T \ \cdots \ \bdomega_{K}^T]^T\right\} \in \Complex{K\Nd}{K\Nd}$.

The channel estimation error associated with $\htbdd_{\rt}$ is defined as $\bde_{\rt} = \bdd_{\rt} - \htbdd_{\rt}$. Note that $\bde_{\rt}$ is distributed as $\bde_{\rt} \sim \clCN(\bdzro,\bdR_{\bde_\rt})$ where $\bdR_{\bde_\rt} \in \Complex{K\Nd}{K\Nd}$ is given by
\begin{align}
	\bdR_{\bde_\rt} 
	= \Xinv{ \xinv{\sigma^2} \bdR_{\rt} \bdA_\rp^H \bdA_\rp + \bdI} \bdR_{\rt}.
\end{align}
Hence, the MMSE  can be written as
\begin{align}
	J = \trace \left( \Xinv{ \xinv{\sigma^2} \bdR_{\rt} \bdA_\rp^H \bdA_\rp + \bdI} \bdR_{\rt} \right). \label{Wideband_MMSE_UT_k}
\end{align}

In the following proposition, we show the asymptotic MMSE in the regime that $\Np$ is sufficiently large, and the conditions under which the asymptotic MMSE can achieve its minimum. 

\begin{myprop} \label{Proposition_MMSE_lowerbound_psp}
	As $\Np \rightarrow \infty$, the MMSE for the optimal channel estimation satisfies $J \rightarrow J_{\asy} $ with $J_{\asy}$ given by
	\begin{align}
		J_{\asy} =  \sum_{s=1}^S \trace \left( \Xinv{ \frac{P}{\sigma^2} \bdR_{\rt,s} \left( \bdC_{s} \otimes \bdI \right) +  \bdI } \bdR_{\rt,s} \right)\comma\label{MMSE_lowerbound_psp}
	\end{align}
	where $\bdR_{\rt,s} = \diag\{[\bdomega_{i}]_{i\in\clK_s}^{\rR}\}$, $\bdC_{s} = \bdG_s^H \bdG_s$ with $\bdG_s = [\bdg_i]_{i\in\clK_s}^{\rC}$. Moreover, $J_{\asy}$ achieves its minimum
	\begin{align}
		J_{\asy}^{\min} = \sum_{k=1}^K \sum_{\ell=0}^{\Nd-1} \frac{\sigma^2 \omega_{k,\ell}}{P  \omega_{k,\ell} + \sigma^2}\comma
	\end{align}
	under the following conditions
	\begin{equation}\label{inter_intra_pilot_interference_null}
		\bdg_i^H \bdg_k = 0\comma\ \forall i,k\in\clK_s\comma\ i \ne k\comma \ \forall s.
	\end{equation}
\end{myprop}
\begin{IEEEproof}
	Please refer to \Cref{appendix_Proposition_MMSE_lowerbound_psp_proof}.
\end{IEEEproof}

In \Cref{Proposition_MMSE_lowerbound_psp}, the asymptotic MMSE $J_{\asy}$ can be written as the sum of $S$ terms, in which the $s$th term is only dependent on the channel parameters of the UTs that use the $s$th pilot. This means that the UTs that use the $s$th pilot would not interfere with other UTs that use different pilots.
Therefore, if $\Np$ is sufficiently large, the interference among the UTs that use different pilots can be completely eliminated, and only the co-pilot interference remains.
In other words, the pilots in $\clX_{\rp}$ tend to be phase shift orthogonal in this case.
In addition, if  the same pilot is used by the UTs whose array response vectors are orthogonal, the co-pilot interference also disappears, and the asymptotic MMSE achieves its minimum. In a nutshell, \Cref{Proposition_MMSE_lowerbound_psp} actually shows under what conditions the pilot reuse can achieve the asymptotically optimal channel estimation performance.

From the above analysis, it can be seen that the channel estimation performance depends critically on the pilot allocation. 
Consequently, it is imperative to carefully design the pilot allocation strategy, which will be studied in the next subsection. 

\subsection{Pilot Allocation} \label{subsec_pilot_allocation}

In this subsection, we focus on the pilot allocation strategy design for LEO satellite massive MIMO OFDM systems.
The optimal pilot allocation strategy should be able to minimize the MMSE for the channel estimation.  The pilot allocation problem can be formulated as a combinatorial optimization problem, whose optimal solution can be obtained through the exhaustive search among $S^K$ feasible solutions.  
Nevertheless, the complexity of exhaustive search is too high and not practical for real-time applications in LEO SATCOM systems. Therefore, it is of great significance to design an efficient pilot allocation strategy, such that the channel estimation performance can be improved to the most extent.

From \Cref{Proposition_MMSE_lowerbound_psp}, we can see that the same pilot should be used by the UTs whose array response vectors are nearly orthogonal, while the UTs whose array response vectors are highly linearly dependent should be allocated with different pilots.
In other words, the inner products between any two array response vectors of the UTs that  use the same pilot should be as close to zero as possible.
Intuitively, we formulate the pilot allocation problem as follows
	\begin{align}\label{Problem_User_Partition}
		\min_{\clK_s,\forall s} &\ \sum_{s=1}^S \sum_{\substack{i,k\in\clK_s, i < k}} W_{i,k} \notag \\
		\mathrm{s.t.} &\ \cup_{s=1}^S \clK_s = \clK \text{ and } \clK_{s} \cap \clK_{s'} = \emptyset\comma \ \forall s'< s\comma
	\end{align}
where $W_{i,k}$ is defined as $W_{i,k} \triangleq \beta_{i} \beta_{k} \sabs{\bdg_{i}^H \bdg_{k}}^2$.
It is worth noting that in \Cref{Problem_User_Partition}, only the large-scale fading parameters $\{\beta_k|\forall k\}$ and the paired space angles $\{\bdxi_k|\forall k\}$ are exploited at the satellite for the pilot allocation. 

\begin{figure}[!t]
	\centering
	\includegraphics[width=0.26\textwidth]{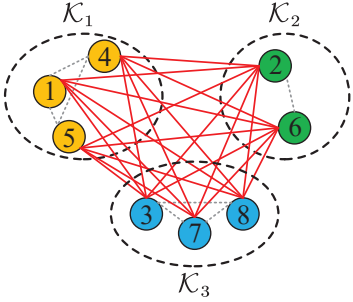}
	\caption{An example of the pilot allocation scheme: $S=3$ pilots are allocated to $8$ UTs by $\clK_1=\{1,4,5\}$, $\clK_2=\{2,6\}$, and $\clK_3 = \{3,7,8\}$.}
	\label{fig_pilot_assignment}
\end{figure}

Next, we show that the pilot allocation problem in \Cref{Problem_User_Partition} can be recast as a max-$S$-cut problem in a graph \cite{Salah2016Scheduling}. 
Let us denote the graph $G(\clK,\clW)$, where $\clK = \{1,\dots,K\}$ represents the vertex set, and  $\clW = \{W_{i,k}|\forall i, k \in \clK\}$ denotes the weight set. Specifically, the edge between each pair of vertices $i$ and $k$ with $i \ne k$, is assigned with a weight $W_{i,k}$, which satisfies $W_{k,i} = W_{i,k}$, and we assume that $W_{i,i} = 0$.
It can be seen that the problem in \Cref{Problem_User_Partition} actually aims to divide the vertex set $\clK$ into $S$ disjoint subsets $\clK_1,\dots,\clK_S$, such that the sum of weights for the edges that connects the vertices within the same subset is minimized. By noticing the following identity
\begin{align}
	& \sum_{s=1}^S \sum_{\substack{i,k\in\clK_s, i < k}} W_{i,k} + \sum_{1\le s'< s \le S} \sum_{i\in\clK_s, i'\in\clK_{s'}} W_{i,i'} \notag \\
	={}& \sum_{1\le i< k\le K} W_{i,k} = \text{constant}\comma
\end{align}
we can derive an equivalent formulation for the problem in \Cref{Problem_User_Partition} as follows
	\begin{align}\label{Problem_Max_S_Cut}
		\max_{\clK_s,\forall s} &\  \sum_{1\le s'< s \le S} \sum_{i\in\clK_s, i'\in\clK_{s'}} W_{i,i'} \notag \\
		\mathrm{s.t.} &\  \cup_{s=1}^S \clK_s = \clK \text{ and } \clK_{s} \cap \clK_{s'} = \emptyset\comma \ \forall s'< s.
	\end{align}
It is worth noting that the problem in \Cref{Problem_Max_S_Cut} is exactly in the form of a max-$S$-cut problem for the graph $G(\clK,\clW)$.
In more detail, the objective of the problem in \Cref{Problem_Max_S_Cut} is to find an $S$-partition of the vertex set $\clK$, such that the sum of weights for the edges whose endpoints are located in different subsets is maximized.

By following the techniques in \cite{Sahni1976Pcomplete}, we present an efficient pilot allocation algorithm in \Cref{algorithm_user_scheduling}, which can achieve a good balance between the computational complexity and channel estimation performance. The computational complexity of \Cref{algorithm_user_scheduling} is $\clO(K^2)$.
In addition, \Cref{algorithm_user_scheduling} is guaranteed to obtain a solution of the problem in \Cref{Problem_Max_S_Cut} with the  achieved objective value no less than $1-\xinv{S}$ times of the maximum value \cite{Sahni1976Pcomplete}.

\begin{algorithm}[!t]
	\caption{Pilot allocation algorithm.} 
	\label{algorithm_user_scheduling}
	\begin{algorithmic}[1]
		\REQUIRE Paired space angles $\{\bdxi_k|\forall k\}$ and average channel power $\{\beta_k|\forall k\}$.
		\ENSURE Pilot allocation results $\clK_1, \dots, \clK_S$.
		\STATE Initialize $\clK_1 = \{1\}$, $\dots$, $\clK_S = \{S\}$, and $\clK^{\mathrm{un}} = \clK\setminus\{1,\dots,S\}$;
		\FOR{$k\in\clK^{\mathrm{un}}$}
		\STATE Compute $s_{k}$ by
		\begin{equation*}
			s_{k}  = \arg \min\limits_{s=1,\dots,S} \left( \sum_{i, i' \in \clK_s, i<i'} W_{i,i'} + \sum_{i\in\clK_s} W_{i,k}  \right)
		\end{equation*}
		\STATE Update $\clK_{s_{k}}: = \clK_{s_{k}} \cup \{k\}$ and $\clK^{\mathrm{un}}: = \clK^{\mathrm{un}}\setminus\{k\}$;
		\ENDFOR
	\end{algorithmic}	
\end{algorithm}

Owing to the existence of the large-dimensional matrix inversion, the computational complexity of the MMSE channel estimation is prohibitively high, which renders it challenging to be implemented in LEO SATCOM systems with the limited satellite payloads. 
The high computational complexity of the MMSE channel estimation arises from the convention that the received signal at the satellite needs to be jointly processed in the space-frequency domain.
However, because the space domain characteristics of the LEO satellite channels can be perfectly known by the satellite, this convention might not be necessary for LEO satellite channel estimation. 
In the next section, we will develop a novel TSCE approach for LEO satellite massive MIMO OFDM communications, which can significantly reduce the computational burden and achieve a near performance as the MMSE estimation. 

\section{Two-Stage Channel Estimation}\label{section_two_stage_channel_estimation}

In this section, we propose a TSCE approach for LEO satellite massive MIMO OFDM systems. First, we show that the MMSE estimator has a two-stage structure in the low SNR and high SNR regimes,  which indicates that the channel estimation can be performed by per-subcarrier space domain processing followed by per-user frequency domain processing. Moreover, the space domain processing of each UT is shown to be identical for all the subcarriers. Motivated by this, we apply the two-stage structure to the general cases. By considering the asymptotic MMSE minimization, an asymptotically optimal vector for the per-subcarrier space domain linear processing is derived. Then, with the help of fast Toeplitz system solver, the per-user frequency domain processing is efficiently implemented to recover the angle-delay domain channel parameters. 
Notice that it is quite challenging for the satellite to acquire the distinct PDP of each UT in LEO SATCOM systems. In the case that the prior knowledge about the PDPs is unavailable, it is reasonable to assume that the UTs have a common PDP according to some typical scenario.
Specifically, we make the following assumption that $\bdR_{\rt,k}$ of each UT $k$ can be written as
\begin{align}\label{Rtk_beta_Gamma}
	\bdR_{\rt,k} = \beta_k \bdGamma\comma
\end{align}
where $\bdGamma = \diag\{\bdgamma\}$ with $ \bdgamma = [\gamma_0\ \cdots\ \gamma_{\Nd-1}]^T$ satisfies $\trace(\bdGamma) = 1$.
This means that the PDPs of all the UTs are assumed to be identical. 
As a consequence, the satellite only needs to know the large-scale fading parameters $\{\beta_k|\forall k\}$ and the common PDP $\bdGamma$ for all the UTs, which can greatly reduce the overhead of statistical CSI acquisition.

\subsection{Analysis for Low SNR and High SNR Regimes} \label{subsec_analysis_lowSNR_highSNR}
First, we show that in the low SNR and high SNR regimes, the MMSE estimator exhibits the two-stage structure. 

\begin{myprop}\label{Proposition_LowSNR_HighSNR}
	For the low SNR regime, which is represented by $\sigma^2 \rightarrow \infty$, the MMSE estimate in \eqref{dfk_estimate_MMSE} can be well approximated by
	\begin{align}
		\htbdd_{\rt} \approx
		\underbrace{\left[\begin{matrix}
				\bdU_{\rL,1}^H & & \\
				& \ddots & \\
				&& \bdU_{\rL,K}^H
			\end{matrix}\right]}_{\udbdU_{\rL}^H} \underbrace{\left[\begin{matrix}
				\bdI_{\Np}\otimes \bdg_1^H \\
				\vdots \\
				\bdI_{\Np} \otimes \bdg_K^H
			\end{matrix}\right]}_{\udbdW_{\rL}^H} \bdy_{\rp}\comma \label{MMSE_estimate_low_SNR}
	\end{align}
	with $\bdU_{\rL,k} = \frac{\sqrt{P/\Np}}{\sigma^2} \FNpNpe \bdD_{\rd,s_k} \bdR_{\rt,k} \in \Complex{\Np}{\Nd}$. 
	In the high SNR regime, i.e., $\sigma^2 \rightarrow 0$, as $\Np\rightarrow \infty$, the MMSE estimate in \eqref{dfk_estimate_MMSE} tends to be reduced to
	\begin{align}
		\htbdd_{\rt} \approx 
		\underbrace{\left[\begin{matrix}
				\bdU_{\rH,1}^H & & \\
				& \ddots & \\
				&& \bdU_{\rH,K}^H
			\end{matrix}\right]}_{\udbdU_{\rH}^H} \underbrace{\left[\begin{matrix}
				\bdI_{\Np} \otimes \bdq_1^H \\
				\vdots \\
				\bdI_{\Np} \otimes \bdq_K^H
			\end{matrix}\right]}_{\udbdW_{\rH}^H} \bdy_{\rp} \comma \label{MMSE_estimate_high_SNR}
	\end{align}
	where $\bdU_{\rH,k} = \xinv{\sqrt{P\Np}} \FNpNpe \bdD_{\rd,s_k} \bdGamma \bdGamma^{\dagger} \in \Complex{\Np}{\Nd}$, $\bdq_k$ is the $i_k$th column vector of $\bdG_{s_k} \bdC_{s_k}^{-1}$ with $i_k$ given by the column index of $\bdg_k$ in $\bdG_{s_k}$.
\end{myprop}
\begin{IEEEproof}
Please refer to \Cref{appendix_Analysis_high_SNR_proof}.
\end{IEEEproof}

\Cref{Proposition_LowSNR_HighSNR} reveals that the channel estimation results in \eqref{MMSE_estimate_low_SNR} and \eqref{MMSE_estimate_high_SNR} exhibit the two-stage structure.  Specifically, the received signals at the satellite are first processed subcarrier-by-subcarrier in the space domain by using $\udbdW_{\rL}$ (or $\udbdW_{\rH}$), and then handled user-by-user in the frequency domain by using $\udbdU_{\rL}$ (or $\udbdU_{\rH}$). Furthermore, each UT's space domain processing is identical for all the subcarriers. Consequently, it is no longer required to jointly process the received signals in the space-frequency domain for LEO satellite channel estimation, in both low SNR and high SNR regimes.

Motivated by the above analysis, we aim to apply the two-stage structure to the general cases in LEO satellite massive MIMO OFDM communications, which will be stated in the next subsection.

\subsection{Two Stage Channel Estimation Approach}
We consider that the channel estimation can be implemented with the following structure
\begin{align}
	\htbdd_{\rt}^{\tsce} = \underbrace{\left[\begin{matrix}
		\bdU_1^H & & \\
		& \ddots & \\
		&& \bdU_K^H
	\end{matrix}\right]}_{\udbdU^H}\underbrace{\left[\begin{matrix}
	\bdI_{\Np} \otimes \bdw_1^H \\
	\vdots \\
	\bdI_{\Np} \otimes \bdw_K^H
\end{matrix}\right]}_{\udbdW^H} \bdy_{\rp}\comma
\end{align}
where $\bdw_k \in \Complex{M}{1}$ and $\bdU_k \in \Complex{\Np}{\Nd}$ represent the per-subcarrier space domain processing vector and per-user frequency domain processing matrix of UT $k$, respectively.
At the first stage, the received signals at the satellite are processed in the space domain with $\bdw_k$ as follows
\begin{align} \label{ywk_two_stage}
	\bdy_{\rw,k} ={}& (\bdI_{\Np} \otimes \bdw_k^H) \bdy_{\rp} 
	= \bdX_{\rc}^* \bdY_{\rp}^T \bdw_k^* \notag \\
	={}& \sqrt{\frac{P}{\Np}} \bdw_k^H \bdg_k \FNpNpe \bdD_{\rd,s_k} \bdd_{\rt,k} \notag \\
	 &\quad + \sum_{i\ne k} \sqrt{\frac{P}{\Np}} \bdw_k^H \bdg_i \FNpNpe \bdD_{\rd,s_i} \bdd_{\rt,i} + \bdz_{\rw,k}\comma
\end{align}
where $\bdz_{\rw,k} = \bdX_{\rc}^* \bdZ_{\rp}^T \bdw_k^*$ is distributed as $\bdz_{\rw,k} \sim \clCN(\bdzro,\sigma^2 \snorm{\bdw_k}^2 \bdI)$. 
Then, at the second stage, the estimate of $\bdd_{\rt,k}$ can be obtained by the per-user frequency domain processing as follows
\begin{align} \label{MMSE_estimation_tsce}
	\htbdd_{\rt,k}^{\tsce} = \bdU_k^H \bdy_{\rw,k}.
\end{align}
Then, with the aim to minimize the MSE $\bbE\{ \snorm{  \bdU_k^H \bdy_{\rw,k} - \bdd_{\rt,k} }^2 \}$, the optimal $\bdU_k$ for the frequency domain processing of UT $k$ should be given by
\begin{align} \label{peruser_freq_domain_process_matrix}
	\bdU_k = \sqrt{\frac{P}{\Np}} \bdw_k^H  \bdg_k \bdT_{\rw,k}^{-1} \FNpNpe \bdD_{\rd,s_k} \bdR_{\rt,k}\comma
\end{align}
where $\bdT_{\rw,k} = \FNpNpe \bdLambda_{\rw,k} \FNpNpe^H\in \Complex{\Np}{\Np}$ 
with the diagonal matrix $\bdLambda_{\rw,k} \in \Real{\Npe}{\Npe}$ given by
\begin{align}
	\bdLambda_{\rw,k} = \sum_{i=1}^K \frac{P}{\Np} \sabs{\bdw_k^H \bdg_i}^2 \bdD_{\rd,s_i} \bdR_{\rt,i} \bdD_{\rd,s_i}^T  + \frac{\sigma^2 \snorm{\bdw_k}^2}{\Npe} \bdI.
\end{align}

The MMSE matrix is given by
\begin{align}
\bdR_{\bde_{\rt},k}^{\tsce} 
= \bdR_{\rt,k} - \frac{P}{\Np} \sabs{\bdw_k^H \bdg_k}^2 \bdV_{\rw,k}\comma
\end{align}
where $\bdV_{\rw,k} = \bdR_{\rt,k} \bdD_{\rd,s_k}^T \FNpNpe^H \bdT_{\rw,k}^{-1} \FNpNpe \bdD_{\rd,s_k} \bdR_{\rt,k}$.
The corresponding MMSE is given by
\begin{align}
	J_{\rw,k} = \trace \left( \bdR_{\rt,k} \right) - \frac{P}{\Np} \sabs{\bdw_k^H \bdg_k}^2 \trace\left( \bdV_{\rw,k} \right).
\end{align}

In the next two subsections, we discuss in details the per-subcarrier space domain processing and per-user frequency main processing, respectively.

\subsection{Per-Subcarrier Space Domain Processing}

In the following proposition, by analyzing the asymptotic property of $J_{\rw,k}$, we derive an asymptotically optimal space domain processing vector $\bdw_k$ that minimizes $J_{\rw,k}$.
\begin{myprop}\label{Proposition_space_domain_combiner_psp}
As $\Np\rightarrow \infty$, the MMSE $J_{\rw,k}$ of UT $k$ satisfies $J_{\rw,k} \rightarrow J_{\rw,k}^{\asy}$. Here, $J_{\rw,k}^{\asy}$ is given by
\begin{align}\label{Jwk_asy}
	J_{\rw,k}^{\asy} = \beta_k - P \beta_k^2 \sabs{\bdw_k^H \bdg_k}^2 \sum_{\ell=0}^{\Nd-1} \frac{\gamma_{\ell}^2}{A_{k,\ell}}\comma
\end{align} 
where $A_{k,\ell} = G_k \gamma_{\ell} + \sigma^2 W_k$, $G_k = P \bdw_k^H \bdG_{s_k} \bdOmega_{s_k} \bdG_{s_k}^H \bdw_k$, $W_k = \snorm{\bdw_k}^2$, and $\bdOmega_{s} = \diag\{[\beta_i]_{i\in\clK_s}^{\rR}\}$.
In addition, the optimal space domain processing vector $\bdw_k$ that minimizes $J_{\rw,k}^{\asy}$ is given by
\begin{align} \label{optimal_wk}
	\bdw_k = \Xinv{ P \bdG_{s_k} \bdOmega_{s_k} \bdG_{s_k}^H + v_k \bdI } \bdg_k\comma
\end{align}
with $v_k$ given by
\begin{align} \label{optimal_vk}
	v_k = \frac{\sum_{\ell=0}^{\Nd-1} \frac{\gamma_{\ell}^2}{A_{k,\ell}^2}}{\sum_{\ell=0}^{\Nd-1} \frac{\gamma_{\ell}^3}{A_{k,\ell}^2}} \sigma^2\comma
\end{align}
which is lowered bounded by $v_k \ge \sigma^2 $ and upper bounded by $v_k \le \frac{\sigma^2}{\bar{\gamma}}$, with $\bar{\gamma}$ given by the average value of the nonzero elements in $\{\gamma_{\ell}|\forall \ell\}$.
\end{myprop}
\begin{IEEEproof}
Please refer to \Cref{appendix_Proposition_space_domain_combiner_psp_proof}.
\end{IEEEproof}

From \Cref{Proposition_space_domain_combiner_psp}, we can see that the asymptotically optimal space domain processing vector $\bdw_k$ of  each UT $k$ should have the form of the regularized zero-forcing (RZF) processing. In addition, $\bdw_k$ only depends on the channel parameters of the UTs that use the same pilot as UT $k$.
Notice that it is generally difficult to obtain the closed form of $v_k$ from \eqref{optimal_vk}, since it is intertwined with $A_{k,\ell}$ that depends on $\bdw_k$ as shown in \Cref{Jwk_asy,optimal_wk}. In addition, we derive the lower bound and upper bound of $v_k$. The conditions to achieve the lower bound and upper bound of $v_k$ are also presented in the following.

The lower bound and upper bound of $v_k$ can become tight under some special channel conditions.
If the LEO satellite channel of each UT only has a single tap, e.g., $\gamma_{\ell'} = 1$ if $\ell = \ell'$, and $\gamma_{\ell} =0$ otherwise,  it can be derived that $v_{k} = \sigma^2$ by using \eqref{optimal_vk}.
On the other hand, if the nonzero elements in $\{\gamma_{\ell}|\forall \ell\}$ are identical, e.g., $\gamma_{\ell} = \bar{\gamma}$ for the nonzero $\gamma_{\ell}$'s, $v_k$ is given by $v_k = \frac{\sigma^2}{\bar{\gamma}}$ due to \eqref{optimal_vk} as well.

According to \eqref{optimal_vk}, $v_k$ is exactly equal to the solution to the nonlinear equation $F(v_k) = 0$, where $F(v_k)$ is given by
\begin{align}
	F(v_k) = \frac{\sum_{\ell=0}^{\Nd-1} \frac{\gamma_{\ell}^2}{A_{k,\ell}^2}}{\sum_{\ell=0}^{\Nd-1} \frac{\gamma_{\ell}^3}{A_{k,\ell}^2}} \sigma^2 - v_k.
\end{align}
In this paper, the Newton's method \cite{Sauer2012NumericalAnalysis} is resorted to solve the nonlinear equation $F(v_k) = 0$, and the detailed procedure is omitted here for conciseness.

\subsection{Per-User Frequency Domain Processing}
From the per-user frequency domain processing in \eqref{peruser_freq_domain_process_matrix}, it can be seen that there exists the term $\bdT_{\rw,k}^{-1} \bdy_{\rw,k}$. By noticing that $\bdT_{\rw,k}$ is a Toeplitz matrix, we only need to focus on the solution to the Toeplitz system $\bdT_{\rw,k} \bdx_{\rw,k} = \bdy_{\rw,k}$.
In this paper, we adopt the classic Levinson recursive algorithm to solve the Toeplitz system \cite{Golub2013MatrixComputations}. For an $n$-dimensional Toeplitz matrix, the number of multiplications required by the Levinson algorithm is  $4n^2$ \cite{Golub2013MatrixComputations}.
We take the example of solving the Toeplitz system $\bdT_n \bds = \bdb$, where $\bdT_{n} \in \Complex{n}{n}$ is a positive definite Toeplitz matrix and $\bdb = [b_1 \ \cdots \ b_{n}]^T$ is an arbitrary vector.
The Levinson algorithm is encapsulated as follows.

First, we rewrite $\bdT_{n}$ as $\bdT_{n} = T_{0} \bdL_{n}$,
where $T_{0} = [\bdT_{n}]_{0,0}$. Then, $\bdL_{n}$ can be written as
\begin{align}
	\bdL_{n} = \left[\begin{matrix}
		1 & \bdr_{n-1}^H \\
		\bdr_{n-1} & \bdL_{n-1}
	\end{matrix}\right]\comma
\end{align}
where $\bdr_{n-1} = [\rho_1\ \cdots \ \rho_{n-1}]^T$.
For convenience, let us denote $\bdr_m = [\rho_1 \ \cdots \ \rho_m]^T$.
The initial values of $\alpha_m$, $\zeta_m$, $\bdx_m \in \Complex{m}{1}$ and $\bdy_m \in \Complex{m}{1}$ are given by $\alpha_1 = - \rho_1$, $\zeta_1 = 1$, $\bdx_1 = b_1$ and $\bdy_1 = - \rho_1$, respectively. 
For $1\le m \le n-1$, $\zeta_{m}$ and $\bdx_m$ are updated as follows
\begin{subequations}\label{Toeplitz_solver}
\begin{align}
	\zeta_{m+1} &= (1-\abs{\alpha_{m}}^2) \zeta_{m}\comma \\
	\mu_{m+1} &= ( b_{m+1} - \bdr_m^T \tdbdx_m ) / \zeta_{m+1}\comma \\
	\bdx_{m+1} &= \left[\begin{matrix}
		\bdx_m + \mu_{m+1} \tdbdy_m^* \\
		 \mu_{m+1}
	\end{matrix}\right]\comma
\end{align}
\end{subequations}
respectively, where $\tdbdx$ denotes the vector with the elements in $\bdx$ rearranged in reverse order. In addition, for $1\le m\le n-2$, $\alpha_m$ and $\bdy_m$ in \eqref{Toeplitz_solver} are updated by
\begin{subequations}\label{Toeplitz_solver_YuleWalker}
	\begin{align}
	\alpha_{m+1} &= -(\rho_{m+1} + \bdr_m^T \tdbdy_m)/\zeta_{m+1}\comma \\
	\bdy_{m+1} &= \left[\begin{matrix}
		\bdy_m + \alpha_{m+1} \tdbdy_m^* \\
		\alpha_{m+1}
	\end{matrix}\right]\comma
\end{align}
\end{subequations}
respectively. After $\bdx_n$ is obtained, the solution to the Toeplitz system $\bdT_n \bds = \bdb$ is given by $\bds = T_0^{-1}\bdx_n$.

Then, once the solution $\bdx_{\rw,k} = \bdT_{\rw,k}^{-1} \bdy_{\rw,k}$ is derived, $\htbdd_{\rt,k}^{\tsce}$ in \eqref{MMSE_estimation_tsce} can be obtained by
\begin{align} \label{MMSE_estimation_tsce_ifft}
	\htbdd_{\rt,k}^{\tsce} &= \sqrt{\frac{P}{\Np}} \bdg_k^H \bdw_k \bdR_{\rt,k} \bdD_{\rd,s_k}^T \FNpNpe^H \bdx_{\rw,k} \notag \\
	&=C_k \bdR_{\rt,k} \bdD_{\rd,s_k}^T \IFFT \left(\left[\begin{matrix}
		\bdzro_{r_\rp\times 1} \\ \bdx_{\rw,k} \\ \bdzro_{R_{\rp}\times 1}
	\end{matrix}\right]\right)\comma
\end{align}
where $C_k =  \sqrt{P\Np} \mud \bdg_k^H \bdw_k $, and $\IFFT(\cdot)$ denotes the inverse fast Fourier transform (IFFT) operation.
The overall description of the TSCE algorithm is summarized in \Cref{algorithm_two_stage}. 

\begin{algorithm}[!t]
	\caption{TSCE algorithm.} 
	\label{algorithm_two_stage}
	\begin{algorithmic}[1]
		\REQUIRE Paired space angles $\{\bdxi_k|\forall k\}$, average channel power $\{\beta_k|\forall k\}$ and common PDP $\{\gamma_{\ell}|\forall \ell\}$.
		\ENSURE Channel estimation results $\htbdd_{\rt,k}^{\tsce}$, $\forall k$.
		\FOR{each UT $k\in\clK$}
		\STATE Compute $\bdw_k$ with \eqref{optimal_wk};
		\STATE Compute $\bdy_{\rw,k} = \bdX_{\rc}^* \bdY_{\rp}^T \bdw_k^*$;
		\STATE Compute $\bdx_{\rw,k} = \bdT_{\rw,k}^{-1} \bdy_{\rw,k}$ with \Cref{Toeplitz_solver,Toeplitz_solver_YuleWalker};
		\STATE Compute $\htbdd_{\rt,k}^{\tsce}$ with \eqref{MMSE_estimation_tsce_ifft};
		\ENDFOR
	\end{algorithmic}	
\end{algorithm}

Then, we discuss the complexity of the MMSE estimation and the TSCE approach.
For the MMSE estimation in \eqref{dfk_estimate_MMSE}, the complexity to compute $\bdA_\rp^H \bdy_{\rp}$ is $\clO(KM\Np + K \Npe \log_2\Npe)$, while the complexity for the matrix inversion is $\clO(K^3\Nd^3)$. Consequently, the total complexity for the MMSE estimation is $\clO(KM\Np + K \Npe \log_2\Npe + K^3\Nd^3)$.
On the other hand, for the TSCE approach in \Cref{algorithm_two_stage}, the complexity to compute  $\{\bdy_{\rw,k}|\forall k\}$ is $\clO(S M^3 + K M^2 + K M\Np)$, and the complexity to compute $\{\htbdd_{\rt,k}^{\tsce}|\forall k\}$ with \eqref{MMSE_estimation_tsce_ifft} is $\clO(K\Np^2 + K\Npe \log_2 \Npe)$.
Therefore, the total complexity of \Cref{algorithm_two_stage} is $\clO( SM^3 + KM^2 + KM\Np + K\Np^2 + K \Npe \log_2 \Npe)$.

\section{Simulation Results} \label{section_simulation_results}
In this section, we show the simulation results to demonstrate the performance of the proposed channel estimation approach in a massive MIMO OFDM LEO SATCOM system. 
The simulation parameters are summarized in \Cref{table_simulation}.
The paired space angles $\{\bdxi_k|\forall k\}$ are generated according to the uniform distribution in the circle region $\{(x,y): x^2 + y^2 \le \sin^2\vtheta_{\max}\}$. 
The elevation angle of UT $k$ is computed by $\alpha_k = \cos^{-1} \left( (R_s/R_e) \sin \vtheta_k \right)$ \cite{Lutz2000SatSysPerson}, where $R_e$ is the earth radius, $R_s = R_e + H$ is the orbit radius.
The distance between the satellite and UT $k$ is given by $D_k = \sqrt{R_e^2 \sin^2 \alpha_k + H^2 + 2 H R_e} - R_e \sin \alpha_k$ \cite{3GPP_NonTerrestrial}.
The per-antenna gains at the satellite and each UT are denoted by $G_{\sat}$ and $G_{\ut}$, respectively.
The propagation delay and power intensity of each path for each UT's channel are generated according to the exponential power delay spectrum.
Moreover, the pathloss and shadow fading of UTs with different elevation angles are computed in accordance with the dense urban scenarios in \cite{3GPP_NonTerrestrial}, while the ionospheric loss is set as $2$ dB approximately.
The OFDM parameters in \Cref{table_simulation} are based on the 5G NR \cite[Table 7.1]{Dahlman20185GNR}.
The noise variance is given by $\sigma^2 = k_\rB T_\rn B / \Nc$ where $k_\rB = 1.38 \times 10^{-23} \text{ J} \cdot \text{K}^{-1}$ is the Boltzmann constant, $T_{\rn}$ is the noise temperature and $B$ is the system bandwidth. 

\begin{table}[!t] 
	\centering
	\footnotesize
	\captionof{table}{Simulation Parameters}
	\label{table_simulation}
	\begin{tabular}{Lc}
		\toprule
		Parameters & Values \\
		\midrule
		Earth radius $R_e$ & $6378$ km \\
		Orbit altitude $H$ & $1000$ km \\
		Central frequency $f_c$  & $2$ GHz \\
		Bandwidth $B$ & $20$ MHz \\
		Noise temperature $T_\rn$ & $290$ K \\
		Number of antennas $\Mx$, $\My$ & $12$, $12$ \\
		Antenna spacing $d_{\rx}$, $d_{\ry}$  & $\lambda$, $\lambda$ \\
		Per-Antenna gain $G_{\sat}$, $G_{\ut}$ & $7$ dBi, $0$ dBi \\
		Maximum nadir angle $\vtheta_{\max}$ & $\xdeg{30}$ \\
		Number of UTs $K$ & $500$ \\
		Transmit power per UT $P$ & $0$ dBW -- $20$ dBW \\
		Subcarrier spacing $\Delta f$  & $60$ kHz \\
		Sampling time interval $\Ts$ & $32.6$ ns \\
		Total number of subcarriers $\Nc$ & $512$ \\
		Number of subcarriers for pilots $\Np$ & $128$ \\
		CP interval $\Ng$ & $36$ \\
		\bottomrule
	\end{tabular}
\end{table}

In this paper, we evaluate the performance by averaging the normalized mean square error (NMSE) of channel estimation over $N_L$ channel samples as follows
\begin{align}
	\mathrm{Averaged\ NMSE} 
	&= \xinv{N_L} \sum_{n=1}^{N_L}  \frac{\sum_{k=1}^K\snorm{\bdH_{\rp,k}(n) - \htbdH_{\rp,k}(n)}_{\mathrm{F}}^2}{\sum_{k=1}^K\snorm{\bdH_{\rp,k}(n)}_{\mathrm{F}}^2} \notag \\
	&=\xinv{N_L} \sum_{n=1}^{N_L} \frac{\snorm{\bdd_{\rp}(n) - \htbdd_{\rp}(n)}^2}{\snorm{\bdd_{\rp}(n)}^2}\comma
\end{align}
where $\bdH_{\rp,k}(n) = \bdg_k (\bdd_{\rp,k}(n))^T$ is the $n$th sample of SFC of UT $k$, $\htbdH_{\rp,k}(n) = \bdg_k (\htbdd_{\rp,k}(n))^T$ is the estimate of $\bdH_{\rp,k}(n)$, $\bdd_{\rp}(n) = [(\bdd_{\rp,1}(n))^T \ \cdots \ (\bdd_{\rp,K}(n))^T]^T$, $\htbdd_{\rp}(n)$ is the estimate of $\bdd_{\rp}(n)$, and $N_L$ is set as $N_L = 100$.

\begin{figure}[!t]
	\centering
	\includegraphics[width=0.49\textwidth]{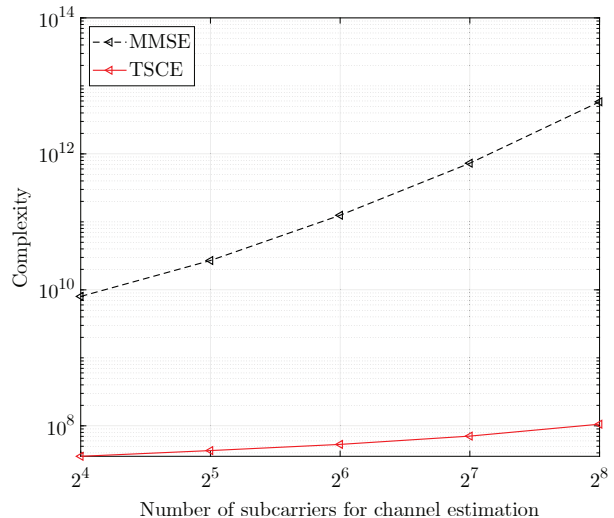}
	\caption{Complexity comparison for the MMSE estimation and the TSCE approach.}
	\label{fig_complexity}
\end{figure}

In \Cref{fig_complexity}, the complexity of the MMSE estimation and the proposed TSCE approach is compared for different values of $\Np$. It can be seen that by adopting the TSCE approach, the complexity of the channel estimation for LEO satellite massive MIMO OFDM communications can be significantly decreased by orders of magnitude compared with the conventional MMSE estimation. Although the TSCE approach has much lower complexity, the simulation results show that its performance can be close to that of the MMSE estimation. 

\begin{figure}[!t]
		\centering
		\includegraphics[width=0.49\textwidth]{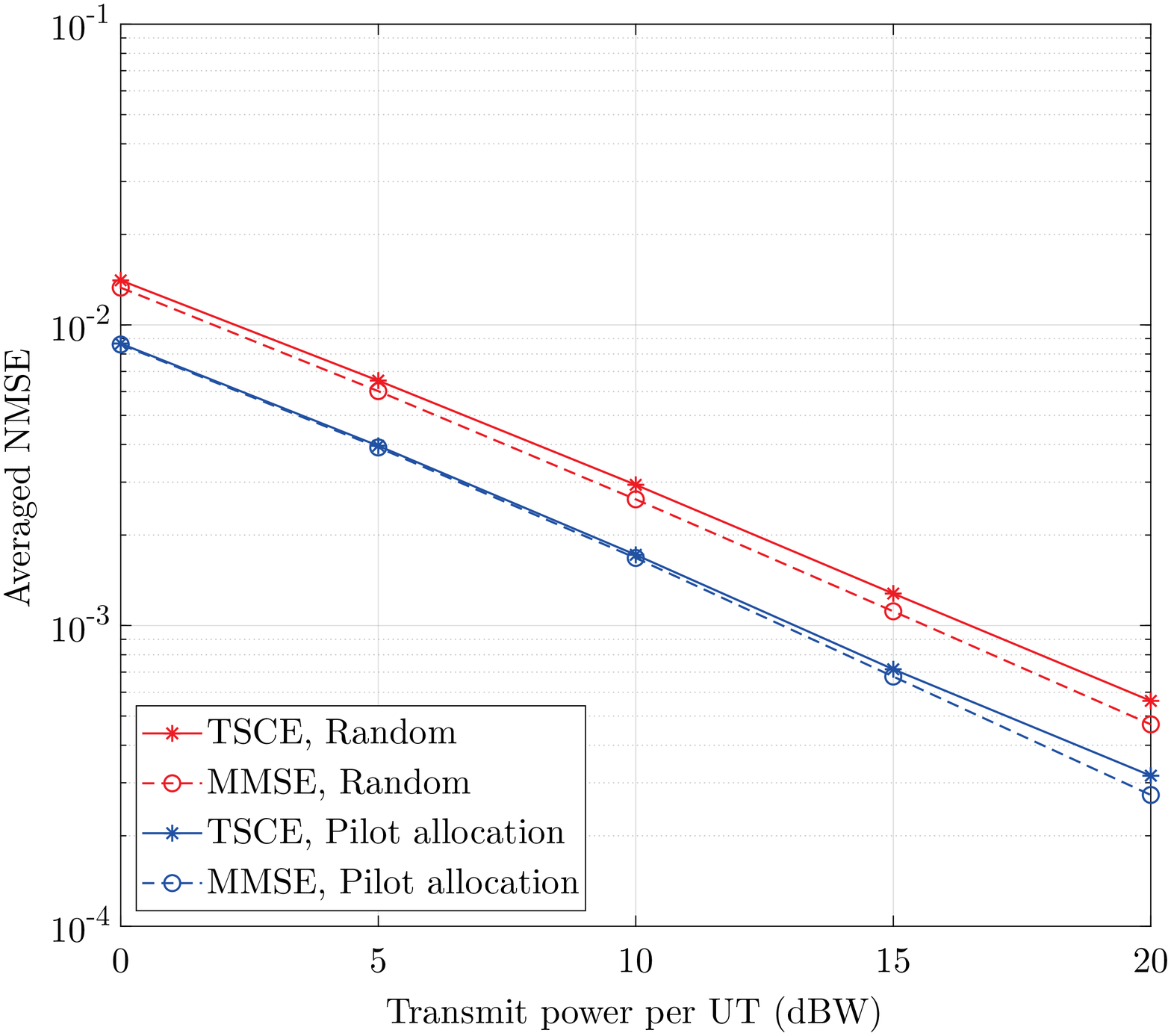}
	\caption{NMSE performance of \Cref{algorithm_user_scheduling}.}
	\label{fig_NMSE_SNR_pilot_allocation}
\end{figure}

\Cref{fig_NMSE_SNR_pilot_allocation} shows the NMSE performance of the proposed pilot allocation strategy in \Cref{algorithm_user_scheduling} and that of the random pilot allocation strategy with $\mud = 2$. It can be observed that, the proposed pilot allocation strategy in \Cref{algorithm_user_scheduling} is capable of significantly improving the channel estimation performance in LEO satellite massive MIMO OFDM communications. In addition, we can see that the TSCE approach can achieve close performance to that of the MMSE estimation, but with much lower computational complexity.

\begin{figure}[!t]
		\centering
		\includegraphics[width=0.49\textwidth]{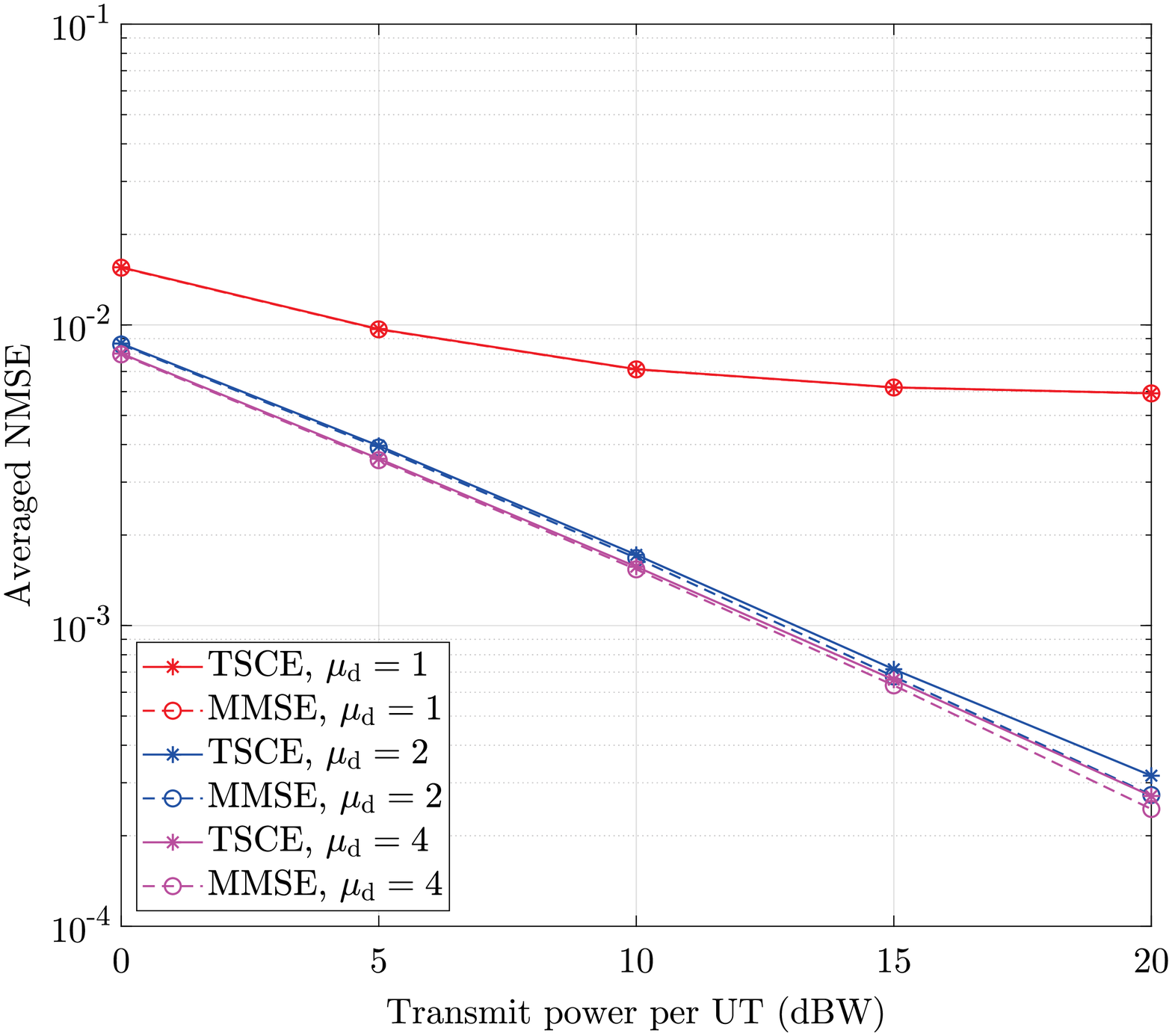}
	\caption{NMSE performance of \Cref{algorithm_two_stage} for different $\mud$.}
	\label{fig_NMSE_SNR_diff_mud}
\end{figure}

\begin{figure}[!t]
	\centering
	\includegraphics[width=0.49\textwidth]{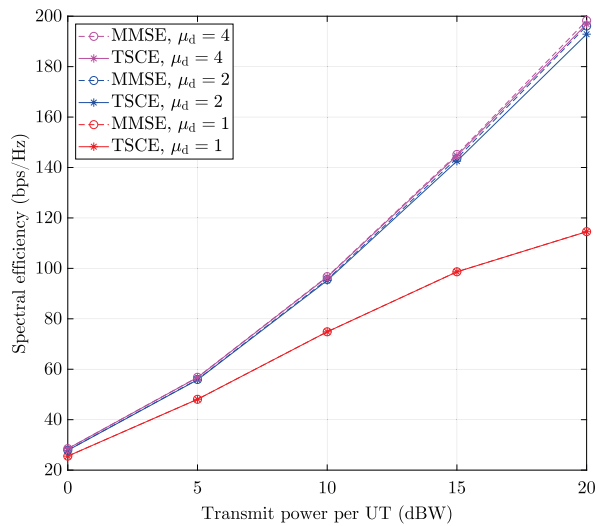}
	\caption{Spectral efficiency performance of Algorithm 2.}
	\label{fig_rate_SNR_diff_Np}
\end{figure}

\begin{figure}[!t]
	\begin{subfigure}{.5\textwidth}
		\centering
		\includegraphics[width=0.999\textwidth]{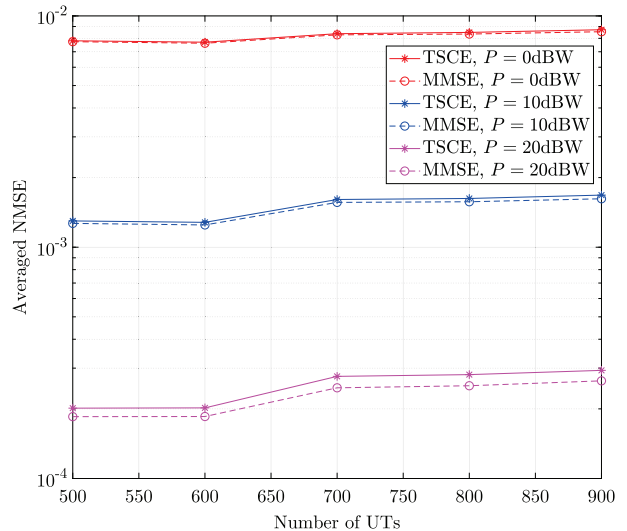}
		\caption{$S=14$.}
		\label{fig_NMSE_SNR_Np_128_diff_UT}
	\end{subfigure}
	\begin{subfigure}{.5\textwidth}
		\centering
		\includegraphics[width=0.999\linewidth]{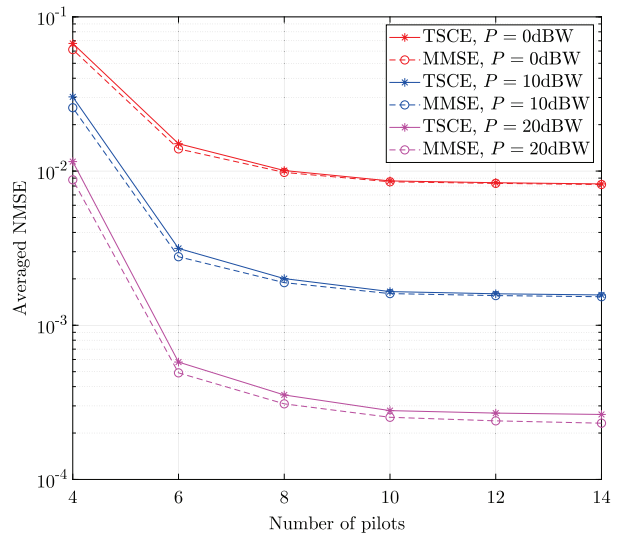}
		\caption{$K=500$.}
		\label{fig_NMSE_SNR_Np_128_diff_S}
	\end{subfigure}
	\caption{NMSE performance of \Cref{algorithm_two_stage} for different $K$ and $S$.}
	\label{fig_NMSE_SNR_diff_UT_S}
\end{figure}

The NMSE performance of the TSCE approach for different values of $\mud$ is shown in Fig. 6. It can be seen that by increasing the value of $\mud$, the channel estimation accuracy can be improved, but at the cost of higher computational complexity. Moreover, the performance improvement brought by increasing $\mud$ from $1$ to $2$ is more significant in contrast to that by increasing $\mud$ from $2$ to $4$. The spectral efficiency of the proposed TSCE approach is illustrated in Fig. 7. We can see that the TSCE approach can also attain near-optimal spectral efficiency performance compared with the MMSE estimation.  Furthermore, the spectral efficiency attained with $\mud=2$ is very close to that with $\mud=4$.

The NMSE performance of the TSCE approach for different values of $K$ is shown in Fig. 8a. It can be seen that the channel estimation accuracy can be well preserved as the number of UTs increases. In addition, Fig. 8b illustrates the NMSE performance of the TSCE approach for different values of  $S$. We can observe that with the increase of the number of pilots, the channel estimation accuracy is obviously improved at first, and then remains at a stable level.
\section{Conclusions} \label{section_conclusion}

In this paper, we have investigated the massive MIMO OFDM channel estimation for LEO SATCOM systems, in which the satellite is equipped with a UPA, and each UT has a single antenna.  
First, we use the ADC to characterize the SFC for LEO satellite massive MIMO OFDM communications. Then, we show that the asymptotic MMSE can be minimized if the array response vectors of the UTs that use the same pilot are orthogonal. Inspired by this, we design an efficient graph-based pilot allocation strategy to enhance the channel estimation performance.
Afterwards, we develop a TSCE approach, in which the channel estimation is carried out by the per-subcarrier space domain processing followed by the per-user frequency domain processing. The space domain processing of each UT is shown to be identical for all the subcarriers, and an asymptotically optimal vector for the per-subcarrier space domain linear processing is derived. By exploiting the fast Toeplitz system solver, the per-user frequency domain processing can be efficiently implemented.
Simulation results verify the near-optimal performance of the proposed channel estimation approach.

\appendices

\section{Proof of \Cref{Proposition_MMSE_lowerbound_psp}}
\label[secinapp]{appendix_Proposition_MMSE_lowerbound_psp_proof}
Notice that the MMSE in  \eqref{Wideband_MMSE_UT_k} can be rewritten as
\begin{align} \label{MMSE_prop_1_proof}
	J = \trace \left( \bdR_{\rt}^{\xinv{2}} \Xinv{ \xinv{\sigma^2} \bdR_{\rt}^{\xinv{2}} \bdA_{\rp}^H \bdA_{\rp} \bdR_{\rt}^{\xinv{2}} +  \bdI } \bdR_{\rt}^{\xinv{2}} \right).
\end{align}
Let us denote $\bdGamma_{i,k} = \bdR_{\rt,i}^{\xinv{2}} \bdD_{\rd,s_i}^T \FNpNpe^H \FNpNpe \bdD_{\rd,s_k} \bdR_{\rt,k}^{\xinv{2}}$. Then, it can be derived that
\begin{align}
	\abs{[\bdGamma_{i,k}]_{\ell,\ell'}} &= \sqrt{\omega_{i,\ell}\omega_{k,\ell'}} \abs{ \sum_{n=0}^{\Np-1} e^{-j\frac{2\pi (r_{\rp}+n)(\phi_{s_k}-\phi_{s_i}+\ell'-\ell)}{\Npe}} } \notag \\
	&= \sqrt{\omega_{i,\ell}\omega_{k,\ell'}} F_{\Np} \left( \frac{ \phi_{s_k}-\phi_{s_i}+\ell'-\ell }{ \mud } \right) \comma
\end{align}
where $F_{\Np}(x) = \abs{\frac{\sin(\pi x)}{\sin(\pi x/\Np)}}$.
As $\Np \rightarrow \infty$, by taking advantage of the rapidly decreasing property of $F_{\Np}(x)$, $\abs{[\bdGamma_{i,k}]_{\ell,\ell'}}$ should tend to satisfy
\begin{align}
	\abs{[\bdGamma_{i,k}]_{\ell,\ell'}} - \Np \sqrt{\omega_{i,\ell}\omega_{k,\ell'}}  \delta(s_k - s_i) \delta(\ell'-\ell) \rightarrow 0.
\end{align}
Then, as $\Np\rightarrow \infty$, we have
\begin{align}
	&\xinv{\sigma^2} \bdR_{\rt}^{\xinv{2}} \bdA_{\rp}^H \bdA_{\rp} \bdR_{\rt}^{\xinv{2}} + \bdI \notag \\
	={}& \frac{P}{\sigma^2\Np} \left[\begin{matrix}
		\bdGamma_{1,1} & \cdots & \bdg_1^H \bdg_K \bdGamma_{1,K} \\
		\vdots & \ddots & \vdots \\
		\bdg_K^H \bdg_1 \bdGamma_{K,1} & \cdots & \bdGamma_{K,K}
	\end{matrix}\right] + \bdI \notag \\
	\rightarrow{}& \frac{P}{\sigma^2} \bdR_{\rt}^{\xinv{2}} (\bdC_{\delta} \otimes \bdI) \bdR_{\rt}^{\xinv{2}} + \bdI\comma
\end{align}
where $\bdC_{\delta}\in\Complex{K}{K}$ with the $(i,k)$th element given by $[\bdC_{\delta}]_{i,k} = \bdg_i^H \bdg_k \delta(s_k - s_i)$.
Accordingly, for the MMSE in \eqref{MMSE_prop_1_proof} we have
\begin{align}
	J &\rightarrow \trace \left( \bdR_{\rt}^{\xinv{2}} \Xinv{ \frac{P}{\sigma^2} \bdR_{\rt}^{\xinv{2}} (\bdC_{\delta} \otimes \bdI) \bdR_{\rt}^{\xinv{2}} + \bdI } \bdR_{\rt}^{\xinv{2}} \right) \notag \\
	&= \sum_{s=1}^S\trace \left( \bdR_{\rt,s}^{\xinv{2}} \Xinv{ \frac{P}{\sigma^2} \bdR_{\rt,s}^{\xinv{2}} \left( \bdC_{s} \otimes \bdI \right) \bdR_{\rt,s}^{\xinv{2}} +  \bdI } \bdR_{\rt,s}^{\xinv{2}} \right) \notag \\
	&= \sum_{s=1}^S \trace \left( \Xinv{ \frac{P}{\sigma^2} \bdR_{\rt,s} \left( \bdC_{s} \otimes \bdI \right) +  \bdI } \bdR_{\rt,s} \right) \notag \\
	&\stackgeq{a} \sum_{s=1}^{S} \sum_{i\in\clK_{s}} \sum_{\ell=0}^{\Nd-1} \frac{\sigma^2\omega_{i,\ell}}{P \omega_{i,\ell} + \sigma^2} \notag \\
	&= \sum_{k=1}^K \sum_{\ell=0}^{\Nd-1} \frac{\sigma^2 \omega_{k,\ell}}{P\omega_{k,\ell} + \sigma^2}\comma \label{MMES_lowerbound_psp_proof}
\end{align}
where $\bdR_{\rt,s} = \diag\{[\bdomega_{i}]_{i\in\clK_s}^{\rR}\}$, $\bdC_{s} = \bdG_s^H \bdG_s$ with $\bdG_s = [\bdg_i]_{i\in\clK_s}^{\rC}$, and (a) follows from the inequality $[\bdA^{-1}]_{i,i} \ge [\bdA]_{i,i}^{-1}$ for positive definite matrix $\bdA$ \cite{Horn2013MatrixAnalysis}.
In addition, the inequality in \eqref{MMES_lowerbound_psp_proof} holds with equality if $\bdC_{s}$ is a diagonal matrix, i.e., $\bdg_i^H \bdg_k = 0$, $\forall i, k\in\clK_s$, $i\ne k$, $\forall s$. This concludes the proof.

\section{Proof of \Cref{Proposition_LowSNR_HighSNR}}
\label[secinapp]{appendix_Analysis_high_SNR_proof}
In case of low SNR, i.e., $\sigma^2 \rightarrow \infty$, it is not difficult to show that
\begin{align}
	\htbdd_{\rt} \approx{}& \xinv{\sigma^2} \bdR_{\rt} \bdA_{\rp}^H \bdy_{\rp} =
	\left[
	\begin{matrix}
		\bdU_{\rL,1}^H \otimes \bdg_1^H \\
		\vdots \\
		\bdU_{\rL,K}^H \otimes \bdg_K^H
	\end{matrix}\right] \bdy_{\rp} \notag \\
	={}&\udbdU_{\rL}^H \udbdW_{\rL}^H \bdy_{\rp}\comma \label{MMSE_estimate_low_SNR_proof}
\end{align}
To show \eqref{MMSE_estimate_high_SNR}, we begin with the derivation of the estimate of $\alpha_{k,\ell}$, $\forall k\in\clK$, $\ell \in \clN_{\rd}$.
According to the proof in \Cref{Proposition_MMSE_lowerbound_psp}, as $\Np \rightarrow \infty$, the estimate of $\alpha_{k,\ell}$ should tend to satisfy
\begin{align}
	\hat{\alpha}_{k,\ell} 
	- \sqrt{\frac{P}{\Np}} ( \bdf_{k,\ell}^H \otimes \bdq_{k,\ell}^H ) \bdy_{\rp}\rightarrow0\comma
\end{align}
where $\bdf_{k,\ell} \in \Complex{\Np}{1}$ is the $\ell$th column vector of $\FNpNpe \bdD_{\rd,s_k}$, $\bdq_{k,\ell} \in \Complex{M}{1}$ is the $i_k$th column vector of $\bdQ_{k,\ell}$ with $\bdQ_{k,\ell} = \gamma_{\ell} \bdG_{s_k} \bdOmega_{s_k} \Xinv{ P \gamma_{\ell} \bdC_{s_k} \bdOmega_{s_k} +  \sigma^2 \bdI }$,
$\bdOmega_{s} = \diag\{[\beta_i]_{i\in\clK_{s}}^{\rR}\}$ and $i_k$ denotes the column index of $\bdg_k$ in $\bdG_{s_k}$.
Then, it can be derived that 
\begin{align}
	\lim_{\sigma^2\rightarrow 0} \bdQ_{k,\ell}
	= \begin{cases}
		\bdzro \comma & \text{ if } \gamma_{\ell} = 0 \\
		\xinv{P} \bdG_{s_k} \bdC_{s_k}^{-1}\comma & \text{ if } \gamma_{\ell} > 0.
	\end{cases}
\end{align}
Then, as $\sigma^2 \rightarrow 0$ and $\Np \rightarrow \infty$, the estimate $\htbdd_{\rt,k} = [\hat{\alpha}_{k,0} \ \cdots \ \hat{\alpha}_{k,\Nd-1}]^T$ of each UT $k$ should satisfy
\begin{align}
	\htbdd_{\rt,k}- \xinv{\sqrt{P\Np}} (\bdGamma \bdGamma^{\dagger} \bdD_{\rd,s_k}^T \FNpNpe^H \otimes \bdq_k^H) \bdy_{\rp} \rightarrow \bdzro.
\end{align}
In other words, for the high SNR cases, as $\Np \rightarrow \infty$, it can be derived that
\begin{align}
	\htbdd_{\rt}\approx \left[\begin{matrix}
		\bdU_{\rH,1}^H \otimes \bdq_1^H \\
		\vdots \\
		\bdU_{\rH,K}^H \otimes \bdq_K^H
	\end{matrix}\right] \bdy_{\rp} = \udbdU_{\rH}^H \udbdW_{\rH}^H \bdy_{\rp}.
\end{align}
This concludes the proof.

\section{Proof of \Cref{Proposition_space_domain_combiner_psp}}
\label[secinapp]{appendix_Proposition_space_domain_combiner_psp_proof}
We begin by rewriting $J_{\rw,k}$ as $J_{\rw,k} = \beta_k - \frac{P}{\Np} \beta_k^2 \sabs{\bdw_k^H \bdg_k}^2 \Gamma_{k}$,
where $\Gamma_{k}$ is given by
\begin{align}
	\Gamma_{k} ={}& \trace \left(\bdGamma \bdF_{\rd,s_k}^H \Xinv{ \bdF_{\rd} \bdR_{\rd,k} \bdF_{\rd}^H + \sigma^2 \snorm{\bdw_k}^2 \bdI } \bdF_{\rd,s_k} \bdGamma \right)\notag \\
	={}& \xinv{\sigma^2 \snorm{\bdw_k}^2} \trace \left(  \bdGamma \bdF_{\rd,s_k}^H \left( \bdI - \bdF_{\rd} \bdR_{\rd,k}^{\xinv{2}} \Xinv{ \sigma^2 \snorm{\bdw_k}^2 \bdI \right.\right. \right.\notag \\
		&\quad \left.\left.\left.+ \bdR_{\rd,k}^{\xinv{2}} \bdF_{\rd}^H \bdF_{\rd} \bdR_{\rd,k}^{\xinv{2}} } \bdR_{\rd,k}^{\xinv{2}} \bdF_{\rd}^H \right) \bdF_{\rd,s_k} \bdGamma \right) \comma
\end{align}
with $\bdF_{\rd,s} = \FNpNpe \bdD_{\rd,s}$, $\bdF_{\rd} = \left[\bdF_{\rd,1} \ \cdots \ \bdF_{\rd,S} \right]$ and $\bdR_{\rd,k}$ is defined as
\begin{align}
	\bdR_{\rd,k} ={}&  \diag\left\{
		\sum_{i\in\clK_{1}} \frac{P}{\Np} \beta_i \sabs{\bdw_k^H \bdg_i}^2,\dots, \right. \notag \\
		&\quad \left. \sum_{i\in\clK_{S}} \frac{P}{\Np} \beta_i \sabs{\bdw_k^H \bdg_i}^2 \right\} \otimes  \bdGamma.
\end{align}
In the light of the proof in \Cref{Proposition_MMSE_lowerbound_psp}, as $\Np\rightarrow \infty$, $\Gamma_k$ tends to be
\begin{align}
	\Gamma_k \approx{}& \trace \left( \bdGamma^2 \left( \sum_{i\in\clK_{s_{k}}} \frac{P}{\Np} \beta_i \sabs{\bdw_k^H \bdg_i}^2 \bdGamma +  \frac{\sigma^2 \snorm{\bdw_k}^2}{\Np} \bdI  \right)^{-1} \right) \notag \\
	={}& \Np \sum_{\ell=0}^{\Nd-1} \frac{\gamma_{\ell}^2}{ A_{k,\ell} }.
\end{align}

Therefore, as $\Np\rightarrow \infty$, it can be derived that $J_{\rw,k} \rightarrow \beta_k - P \beta_k^2 \sabs{\bdw_k^H \bdg_k}^2 \sum_{\ell=0}^{\Nd-1} \frac{\gamma_{\ell}^2}{A_{k,\ell}} = J_{\rw,k}^{\asy}$.
At the optimal point $\bdw_k$, the gradient of $J_{\rw,k}^{\asy}$ with respect to $\bdw_k$ should vanish, i.e., $\nabla_{\bdw_k} J_{\rw,k}^{\asy} = \bdzro$.
As a result, we can obtain that
\begin{align}
	&\bdg_k \cdot \bdg_k^H \bdw_k \sum_{\ell=0}^{\Nd-1} \frac{\gamma_{\ell}^2}{A_{k,\ell}} \notag \\
	={}& \sabs{\bdw_k^H \bdg_k}^2 \sum_{\ell=0}^{\Nd-1} \frac{ \gamma_{\ell}^2}{A_{k,\ell}^2} \left( P \gamma_{\ell} \bdG_{s_k} \bdOmega_{s_k} \bdG_{s_k}^H + \sigma^2 \bdI \right) \bdw_k \notag \\
	={}& A_k \left( P \bdG_{s_k} \bdOmega_{s_k} \bdG_{s_k}^H + v_k \bdI\right) \bdw_k\comma
\end{align}
where $A_k = \sabs{\bdw_k^H \bdg_k}^2 \sum_{\ell=0}^{\Nd-1} \frac{\gamma_{\ell}^3}{A_{k,\ell}^2}$.
By noticing the fact that $\alpha \bdw_k$ can achieve the identical MMSE performance with $\bdw_k$ for any non-zero scalar $\alpha \in \bbC$, the optimal $\bdw_k$ should have the form in \eqref{optimal_wk}.
According to \eqref{optimal_vk}, we have $v_k \ge \left(\sum_{\ell=0}^{\Nd-1} \frac{\gamma_{\ell}^3}{A_{k,\ell}^2}\right)\left(\sum_{\ell=0}^{\Nd-1} \frac{\gamma_{\ell}^3}{A_{k,\ell}^2}\right)^{-1} \sigma^2
=\sigma^2$,
where the inequality comes from $\gamma_{\ell}^2 \ge \gamma_{\ell}^3$, $\forall \ell$, due to $0\le\gamma_{\ell}\le1$ in \eqref{Rtk_beta_Gamma}.
Meanwhile, $v_k$ in \eqref{optimal_vk} can be rewritten as
$v_k = \left(\sum_{\ell=0}^{\Nd-1} \frac{\gamma_{\ell}^2}{A_{k,\ell}^2}\right)\left(\sum_{\ell=0}^{\Nd-1} \frac{\gamma_{\ell}^2}{A_{k,\ell}^2}\gamma_{\ell} \right)^{-1} \sigma^2 
= \left(\sum_{\ell=0}^{\Nd-1} p_{k,\ell} \gamma_{\ell}\right)^{-1} \sigma^2$,
where $p_{k,\ell} = \left(\frac{\gamma_{\ell}^2}{A_{k,\ell}^2}\right)\left(\sum_{\ell'=0}^{\Nd-1} \frac{\gamma_{\ell'}^2}{A_{k,\ell'}^{2}}\right)^{-1}$, $\forall \ell$.
Notice that $\{p_{k,\ell}|\forall \ell\}$ can be viewed as a probability distribution because of $\sum_{\ell=0}^{\Nd-1} p_{k,\ell} = 1$.
From the relation $p_{k,\ell} \ge p_{k,\ell'}$ for $\gamma_{\ell} \ge \gamma_{\ell'}$, we can derive the inequality $\sum_{\ell=0}^{\Nd-1} p_{k,\ell} \gamma_{\ell} \ge \bar{\gamma}$. Therefore, an upper bound of $v_k$ can be obtained by $v_k \le \frac{\sigma^2}{\bar{\gamma}}$.
This concludes the proof.

\bibliographystyle{IEEEtran}  
\bibliography{IEEEabrv,satellite_channel_estimation_lib} 

\end{document}